\documentclass[print,
               notlongauthorlist
              ]{nsr}

\usepackage{tikz}
\usetikzlibrary{arrows.meta,positioning,shapes.geometric}
\usepackage{siunitx}
\newcommand{\dmuon}{$d\mu$}
\newcommand{\tmuon}{$t\mu$}

\volume{00}
\artnum{00}
\firstpage{1}
\datesubmitted{XX Month 2026}
\datereceived{XX Month 2026}
\daterevised{XX Month 2026}
\dateaccepted{XX Month 2026}
\datepublished{XX Month 2026}
\doinum{doi/number}
\copyrightyear{2026}

\author[1]{L. V. Dudko\footnote{Corresponding author e-mail address: \tt dudko@sinp.msu.ru}$^{\orcidlink{0000-0002-4462-3192}}$}

\affil[1]{Skobeltsyn Institute of Nuclear Physics, Lomonosov Moscow State University, 1(2) Leninskie Gory, Moscow 119991, Russian Federation}

\runauth{L. V. Dudko}

\title{The muon collider: expected physics, technological challenges, and the prospects for reusing the 21~km UNK tunnel}

\begin{document}

\maketitle

\begin{abstract}
\noindent
A multi-TeV muon collider is now among the main options considered for the post-LHC energy frontier. The muon is an elementary particle that radiates $\sim\!10^9$ times less synchrotron power than an electron of the same energy. A circular muon collider would therefore deliver the full beam energy to the hard collision in a compact collider ring, although its acceleration chain needs a longer tunnel. It combines the energy reach of a $100$~TeV-class proton machine with the clean final states of a lepton collider. This review summarizes the expected physics results. They include Higgs couplings and self-couplings, electroweak and vector-boson-fusion processes, the top quark, and a broad programme of searches beyond the Standard Model. The programme would close nearly all of the thermal window for electroweak WIMP dark matter. The review also covers the technological challenges, namely muon production, ionization cooling, rapid acceleration, high-field high-temperature-superconductor (HTS) magnets, the machine--detector interface, and the neutrino-flux constraint. For each, the review describes the proposed solution, its present status, and the open issues. None of these technologies has yet been demonstrated at the required performance, and the design luminosity is a target. The review then examines three ways of reusing the existing 21~km UNK tunnel near Protvino, by scaling the parameters of the International Muon Collider Collaboration (IMCC). If the tunnel houses the last rapid-cycling synchrotron, as the LHC tunnel does in the IMCC option at CERN, this synchrotron limits the centre-of-mass energy to about $6$~TeV. A new collider ring of about $7$--$9$~km is then needed. If the tunnel houses the collider ring, arc dipoles of $11$--$14$~T give about $14$--$18$~TeV, and aspirational $16$~T dipoles about $21$~TeV. The acceleration chain then needs a separate tunnel of about $50$--$73$~km. If the tunnel houses both the collider ring and the last synchrotron, as proposed earlier for the LHC tunnel, only the first synchrotrons need new ring tunnels. This layout gives about $6$~TeV with the IMCC baseline magnets and about $10$~TeV only with $16$~T fixed and $\pm3.8$~T pulsed synchrotron dipoles, neither yet demonstrated, at $35$--$57\%$ of the target luminosity. Such a facility would also support a programme of non-collider applications of intense muon beams, such as an aimed neutrino beam, as well as muon-catalyzed fusion, muography, muonic-atom isotopic analysis, and muon spin spectroscopy. Some of these directions have a long history in the Russian physics institutes.
\end{abstract}


{\hypersetup{linkcolor=black}
\tableofcontents}

\section{Introduction}
\label{sec:intro}

The discovery of the Higgs boson in 2012 completed the particle content of the Standard Model (SM). Several basic questions remain open: the nature of electroweak symmetry breaking, the origin of the matter--antimatter asymmetry, the identity of dark matter, and the absence so far of any new state at the Large Hadron Collider (LHC). Answering them calls for a machine that combines high energy reach with measurement precision, two goals that have traditionally belonged to different machines. Hadron colliders reach high energy but collide composite protons, so only a small and smeared fraction of the beam energy enters the elementary collision. For heavy-pair production a $pp$ machine needs a centre-of-mass energy $7$--$20$ times higher than a point-like collider to reach the same masses. The partonic collisions that produce such pairs typically carry only $\sim\!1/10$ of the proton--proton collision energy. Single $s$-channel resonances need a smaller factor, about $5$~\cite{Delahaye2019,MuonSmasher2021}. Lepton colliders collide point-like particles with the full energy available and in a cleaner environment. Electrons and positrons, however, radiate away energy so rapidly in a ring that the practical circular limit was reached at LEP. There each beam radiated about $3\%$ of its energy on every turn, $2.9$~GeV per turn already at a beam energy of $100$~GeV~\cite{LEPEnergyWG2005,PDG2022}. Linear colliders avoid this limit, but the dense bunches radiate in each other's field at the collision point (beamstrahlung). The beamstrahlung energy loss grows rapidly with energy. It is $\approx1\%$ at the $250$~GeV ILC and $4.5\%$ at $500$~GeV~\cite{ILCTDR2013}. At the $3$~TeV CLIC it reaches $28\%$, and only $35\%$ of the luminosity lies within $1\%$ of the nominal energy~\cite{CLICCDR2012}.

A high-energy muon collider avoids this trade-off. The muon is, as far as is known, an elementary particle, so the entire centre-of-mass energy $\sqrt{s}$ is delivered to the hard process, as at an $e^+e^-$ machine. At the same time the muon is $\approx 207$ times heavier than the electron. Because synchrotron radiation scales as $1/m^4$ at fixed energy and radius, a muon radiates roughly $207^4 \approx 1.8\times10^9$ times less power. A muon beam can therefore be accelerated and stored in a compact ring at energies far beyond the reach of any circular $e^+e^-$ collider, and re-used for more than a thousand turns. In particular, at a fixed average bending field $\bar B$ the luminosity lifetime measured in turns does not depend on the beam energy, because the circumference grows in proportion to the energy. With $14$~T arc dipoles ($\bar B\approx9.2$~T) it is $\approx1360$ turns~\cite{MuColParams2025,IMCC2025}. This estimate accounts for muon decay only. Section~\ref{sec:ring} gives the derivation and discusses the other mechanisms of beam loss. Detailed studies show that, for the electroweak pair-production and vector-boson-fusion processes that dominate at these energies, a $10$~TeV muon collider has a physics reach comparable to that of a $\sim\!100$~TeV proton--proton collider. Its collider ring is an order of magnitude shorter than a 100~TeV proton ring~\cite{Delahaye2019,MuonSmasher2021,Long2021NatPhys,Costantini2020,IMCC2023}. The full complex is larger. Including the acceleration chain, the IMCC greenfield design at $10$~TeV needs about $70$~km of tunnels~\cite{IMCC2025} (Section~\ref{subsec:layouts}).

\subsection{The single hard problem and a brief history}

Relative to the electron, the price of these advantages is the muon's instability: at rest it decays with a lifetime of only $\tau_\mu = 2.2~\mu\mathrm{s}$. Every stage of a muon collider (production, cooling, acceleration, and storage) must be completed within this time. This is possible only because relativistic time dilation stretches the laboratory lifetime in proportion to the Lorentz factor $\gamma$. The idea has a long history. Colliding muon beams were first proposed at the Joint Institute for Nuclear Research (JINR) in Dubna by Tikhonin in 1968~\cite{Tikhonin1968}. Budker proposed them independently, emphasizing the radiation advantage, at the 1969 Yerevan accelerator conference~\cite{Budker1969}. Ionization cooling is the only method fast enough to reduce the phase space of a muon beam within its lifetime. It was proposed by Ado and Balbekov at the Institute for High Energy Physics (IHEP) in Protvino in 1971~\cite{AdoBalbekov1971}. Skrinsky and Parkhomchuk proposed it independently at Novosibirsk~\cite{Skrinsky1981}, and Neuffer developed it~\cite{Neuffer1983}. A sustained design effort followed in the United States through the Neutrino Factory and Muon Collider Collaboration~\cite{Feasibility1996,Ankenbrandt1999,Palmer2014}. The international Muon Ionization Cooling Experiment (MICE) observed the ionization-cooling effect, a reduction of the transverse emittance, in a single absorber and in the absence of beam re-acceleration~\cite{MICE2020,MICE2024,IMCC2025}. Cooling of a bunched beam in all six phase-space dimensions, with re-acceleration, is still to be demonstrated (Section~\ref{sec:technology}).

\subsection{Strategic context}

The muon collider is today the focus of a coordinated international programme. The 2020 update of the European Strategy for Particle Physics called for an intensified accelerator R\&D effort, which led directly to the formation of the International Muon Collider Collaboration (IMCC)~\cite{ESPPU2020,AccelRoadmap2022}. The IMCC now includes more than $450$ participants and has produced a comprehensive design report and consolidated parameter sets~\cite{IMCC2023,IMCCInterim2024,IMCC2025,MuColParams2025}. In the United States, the 2023 Particle Physics Project Prioritization Panel (P5) named the muon collider a ``muon shot'' and recommended that the country prepare to host such a machine~\cite{P5Report2023,ForumReport2024}. A US Muon Collider Collaboration was formed in 2024~\cite{USWhitePaper2025,USInput2026}. The first recommendation of the 2025 National Academies report is for the United States to host the world's highest-energy collider by mid-century~\cite{NASEM2025}.

\subsection{Scope of this review}

This article has four main parts. Section~\ref{sec:physics} reviews the expected physics results of a multi-TeV muon collider. Section~\ref{sec:technology} reviews the principal technological challenges and, for each, the proposed solution, its status, and the open issues (Section~\ref{sec:openchallenges}). Section~\ref{sec:unk} is the central part of this review. It assesses quantitatively three ways of reusing the existing 21~km UNK tunnel near Protvino, for the last stage of the acceleration chain, for the collider ring, or for both. Section~\ref{sec:applications} is devoted to the non-collider applications of the intense muon beams that such a facility would develop, led by the neutrino programme. These fields have a long history in the Russian physics institutes at Dubna, Moscow, Novosibirsk, and Protvino. Sections~\ref{sec:discussion} and~\ref{sec:conclusion} discuss the comparative picture and conclude.

\section{Expected physics results}
\label{sec:physics}

\subsection{Energy, luminosity, and the vector-boson-fusion regime}

The muon-collider programme is conceived in energy stages. A first stage at $\sqrt{s}=3$~TeV would demonstrate the technology and start the physics programme. In the next stage, the reference machine operates at $\sqrt{s}=10$~TeV. A $14$~TeV option, together with the higher energies discussed in Section~\ref{sec:unk}, defines the ultimate reach. The consolidated parameter set of the Horizon Europe MuCol project~\cite{MuColParams2025} is summarized in Table~\ref{tab:params}. A guiding design requirement is that the integrated luminosity should scale with energy as $\int\!\mathcal{L}\,dt \propto (\sqrt{s}/10~\mathrm{TeV})^2 \times 10~\mathrm{ab}^{-1}$~\cite{BFW2021,MuColParams2025}. The number of events in the highest-mass final states then remains roughly constant as the energy grows. How the luminosity of a storage ring depends on its field and circumference is discussed in Section~\ref{sec:ring}.

\begin{table}[!hbt]
\centering
\caption{Consolidated muon-collider parameters (MuCol/IMCC reference)~\cite{MuColParams2025,IMCC2023}. Values are indicative design targets. The 3 and 10~TeV stages are the consolidated reference, whereas the 14~TeV column is an indicative upgrade option, obtained by luminosity scaling and not part of the consolidated MuCol parameter set. Dipole fields are peak values. The site-power estimates are from the Snowmass Implementation Task Force~\cite{ITF2022}. The more recent IMCC estimate is lower, $\approx115$ and $\approx190$~MW at $3$ and $10$~TeV~\cite{IMCC2025}.}
\label{tab:params}
\begin{tabular}{lccc}\toprule
Parameter & 3~TeV & 10~TeV & 14~TeV \\
\midrule
Peak $\mathcal{L}$ ($10^{34}$~cm$^{-2}$s$^{-1}$) & 2.1 & 18 & $\sim$35 \\
Integrated $\int\!\mathcal{L}\,dt$ (ab$^{-1}$) & 1 & 10 & 20 \\
Ring circumference (km) & 4.5 & 11.4 & $\sim$14 \\
Arc-dipole peak field (T) & 11 & 14 & 16 \\
Muons per bunch ($10^{12}$) & 2.2 & 1.8 & 1.8 \\
Repetition rate (Hz) & 5 & 5 & 5 \\
Proton driver power (MW) & 2 & 2 & 2--4 \\
Site power (MW) & $\sim$230 & $\sim$300 & $\sim$300 \\
\bottomrule
\end{tabular}
\end{table}

\begin{figure}[!htb]
\centering
\includegraphics[width=0.62\linewidth]{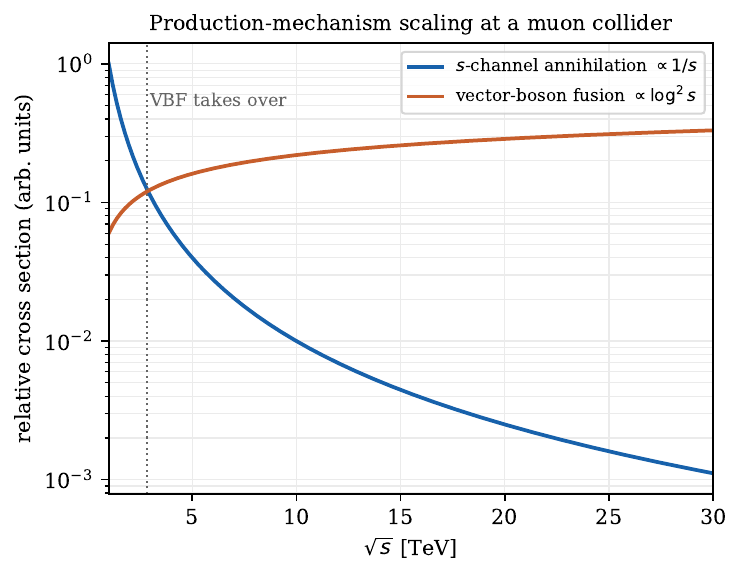}
\caption{Schematic energy dependence of the two production mechanisms at a muon collider: $s$-channel annihilation falls as $1/s$, while vector-boson fusion grows with energy and dominates above a few TeV. The curves are illustrative, and the qualitative crossover follows Ref.~\cite{Costantini2020}.}
\label{fig:xsec}
\end{figure}

At these energies the dominant production mechanisms differ qualitatively from those at lower-energy lepton colliders. Direct $s$-channel annihilation, $\mu^+\mu^-\to X$, has a cross section that falls as $1/s$. The electroweak gauge bosons radiated from the incoming muons, in contrast, act as a dense flux of initial-state partons whose effective luminosity grows logarithmically with energy. Above roughly $1$--$3$~TeV vector-boson fusion (VBF) therefore overtakes annihilation as the dominant production mechanism for most Standard Model final states (Fig.~\ref{fig:xsec})~\cite{Costantini2020}. The same flux of electroweak bosons also drives vector-boson scattering (VBS), $VV\to VV$ with $V=W,Z$, whose longitudinal amplitudes probe the electroweak symmetry-breaking sector directly. A multi-TeV muon collider is thus at the same time a high-energy lepton collider, delivering the full $\sqrt{s}$ to rare $s$-channel processes, and an effective electroweak-boson collider. Much of its Higgs and electroweak programme relies on both roles~\cite{Snowmass2022Summary,EWscattering2024,HanMaXie2021}.

This combination distinguishes the muon collider from the other proposals for the next collider, summarized in Table~\ref{tab:compare}. Circular $e^+e^-$ machines (FCC-ee, CEPC) are precision factories whose energy is capped by synchrotron radiation in the arcs. Linear colliders (ILC, CLIC) avoid this limit. Their energy reach is set by the cost of the accelerating length and by beamstrahlung, which broadens the spectrum of collision energies below the nominal energy (Section~\ref{sec:intro}). The proton collider FCC-hh reaches the highest nominal energy. Its baseline is $\sqrt{s}\approx85$~TeV with $14$~T Nb$_3$Sn dipoles in a $90.7$~km ring and a peak luminosity of $3\times10^{35}$~cm$^{-2}$\,s$^{-1}$. It rests on established hadron-collider technology, although its $14$~T dipoles are still under development. This makes its luminosity projection considerably more mature than that of a muon collider~\cite{FCCFeasibility2025v2}. Its beams are composite, however, so the energy entering a given partonic collision is both smaller and broadly distributed. The muon collider is the only proposal that places a point-like, full-energy collision at the multi-TeV frontier~\cite{MuonSmasher2021,ComparativeEval2025}. Its luminosities above $10^{34}$~cm$^{-2}$\,s$^{-1}$ (Table~\ref{tab:params}), however, need to be demonstrated (Section~\ref{sec:technology}).

\begin{table}[!hbt]
\centering
\caption{The muon collider among the principal proposals for a future high-energy collider. }
\label{tab:compare}
\begin{tabular}{lll>{\raggedright\arraybackslash}p{7.5cm}}\toprule
Machine & Beams & $\sqrt{s}$ & Character / main strength \\
\midrule
FCC-ee, CEPC & $e^+e^-$ (ring) & $0.09$--$0.37$~TeV & Precision Higgs, electroweak and flavour factory, with energy capped by synchrotron radiation \\
ILC, CLIC & $e^+e^-$ (linac) & $0.25$--$3$~TeV & Precision plus moderate energy reach, beamstrahlung-broadened spectrum of collision energies, cost scaling with length \\
FCC-hh & $pp$ (ring) & $85$--$100$~TeV & Highest nominal energy and best reach for coloured states, luminosity based on established hadron-collider technology, but composite beams (lower, spread-out parton energies) \\
Muon collider & $\mu^+\mu^-$ (ring) & $3$--$14$~TeV$^{\,a}$ & Full energy in a point-like collision, precision and energy frontier combined, but $\mathcal{L}>10^{34}$~cm$^{-2}$\,s$^{-1}$ is a design target, not yet demonstrated \\
\bottomrule
\multicolumn{4}{l}{\footnotesize $^a$Up to $\approx21$~TeV with the collider ring in the 21~km UNK tunnel, Section~\ref{sec:unk}.}
\end{tabular}
\end{table}

\subsection{Higgs boson couplings, width, and self-interactions}

The Higgs sector is central to the programme and makes use of both roles of the machine. At $\sqrt{s}=m_H\approx125$~GeV a muon collider can run as an $s$-channel Higgs factory, sitting on the $\mu^+\mu^-\to H$ resonance~\cite{Barger1997}. This should be a separate low-energy collider fed by the cooling complex of the high-energy machine~\cite{IMCC2025}, not the multi-TeV ring itself. In a ring of fixed circumference both the number of turns per fill, Eq.~\eqref{eq:nturns} of Section~\ref{sec:ring}, and the beam density at the collision point fall as the beam energy is lowered. Direct-channel production is feasible only because the muon Yukawa coupling is about $207$ times larger than the electron's. The resonant rate, which scales as the coupling squared, is therefore some $4\times10^4$ times higher than for an $e^+e^-$ machine. The resonance is narrow, $\Gamma_H\approx4.1$~MeV in the Standard Model~\cite{HXSWGYR4}, so the beam-energy spread must be comparable. The benchmark study of Ref.~\cite{deBlasGuLiu2022} assumes a spread of $0.003\%$ per beam, i.e.\ $\sigma_{\sqrt{s}}\approx2.7$~MeV. This energy spread and initial-state radiation reduce the peak cross section of about $70$~pb to an effective on-resonance value of $\approx22$~pb. A line-shape scan around $m_H$ then yields about $7\times10^4$ Higgs bosons for $5$~fb$^{-1}$ and $3\times10^5$ for $20$~fb$^{-1}$~\cite{deBlasGuLiu2022}. These numbers are lower than the on-peak cross section would give, because the scan spends part of the luminosity off the peak.

At multi-TeV energies the high-energy collider produces Higgs bosons copiously through vector-boson fusion. For $10$~ab$^{-1}$ at $\sqrt{s}=10$~TeV (the dataset of one experiment), about $10^7$ single-Higgs events are expected~\cite{IMCC2023,Andreetto2026}. The couplings to vector bosons and to third-generation fermions can then be extracted at the sub-percent to percent level within the $\kappa$ framework~\cite{Forslund2024,Castelli2025}. For $1$~ab$^{-1}$ at $3$~TeV, a first full-simulation study includes the beam-induced background discussed in Section~\ref{sec:technology}. It reaches a statistical precision of $0.75\%$ on $\sigma(H\to b\bar b)$ and a few percent on $H\to WW^*$~\cite{Andreetto2025,Castelli2025}. For $10$~ab$^{-1}$ at $10$~TeV, the corresponding study uses the MUSIC (MUon System for Interesting Collisions) detector concept~\cite{MUSIC2025} and also overlays the incoherent-pair background. It projects statistical precisions of $0.25\%$ on $\sigma(H\to b\bar b)$ and $0.49\%$ on $\sigma(H\to WW^*)$. In the $\gamma\gamma$, $ZZ^*$ and $\mu^+\mu^-$ channels it projects $3.7$--$9.7\%$. The $\gamma\gamma$ and $\mu^+\mu^-$ channels use a parametric simulation tuned to the full one~\cite{Andreetto2026}. The $b\bar b$ measurement is limited by the jet energy scale. Its in-situ calibration to $0.06\%$ translates into a systematic uncertainty of $0.49\%$ on the $b\bar b$ signal yield, about twice the statistical one~\cite{Andreetto2026}.

Two measurements are distinctive for a muon collider. First, the Higgs coupling to the muon itself, $y_\mu$, enters production directly, through $s$-channel production and through $\mu^+\mu^-\to H\gamma$, whose cross section grows with energy~\cite{HiggsMuon2024,Buttazzo2021}. Other machines measure it through the decay $H\to\mu^+\mu^-$. The LHC experiments have evidence for this decay at the $3.0\sigma$ (CMS) and $3.4\sigma$ (ATLAS) level~\cite{CMSHmumu2021,ATLASHmumu2025}. The HL-LHC, FCC-ee and the combined FCC-ee and FCC-hh programme are projected to reach $\approx4\%$, $\approx3\%$ and $\approx0.4\%$ on $\kappa_\mu$~\cite{HLLHCHiggsYR2019,FCCFeasibility2025}, in all cases through a branching ratio and hence together with the total width. For $10$~ab$^{-1}$ at $10$~TeV, a muon collider reaches $\approx2.4\%$ on $\kappa_\mu$ in a global fit including the HL-LHC~\cite{IMCC2025}. For the same luminosity it reaches, independently of the width, $5$--$10\%$ (95\% C.L.) on the $\mu\mu H$ vertex from multi-boson production~\cite{HiggsMuon2024}. A $125$~GeV $s$-channel collider with $5$--$20$~fb$^{-1}$ would measure the muon coupling to about half a percent and $\Gamma_H$ from the line shape to a few percent~\cite{deBlasGuLiu2022,IMCC2025}. Second, the total Higgs width can be determined without model assumptions. The degeneracy between the inclusive Higgs production rate and the total width is broken by tagging the forward muons of the $ZZ$-fusion channel. With a dedicated forward-muon detector covering $|\eta|<6$ and $10$~ab$^{-1}$ at $10$~TeV, the inclusive Higgs rate is measured to $0.75\%$, and the total width is then pinned to the interval $(-0.41\%,+2.1\%)$~\cite{Li2024forward,ForwardMuon2024}. A comparable model-independent width determination ($\sim\!1\%$) is available at FCC-ee through the $ZH$ recoil method~\cite{FCCFeasibility2025}. At a muon collider it comes from the same machine that probes the multi-TeV frontier.

\begin{figure}[!htb]
\centering
\includegraphics[width=0.6\linewidth]{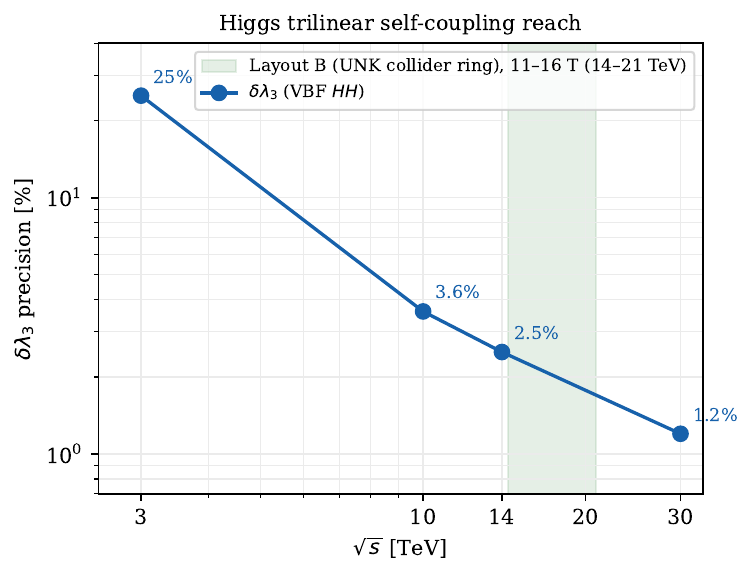}
\caption{Projected precision on the trilinear Higgs self-coupling $\lambda_3$ as a function of $\sqrt{s}$, from the energy-dependent reach of Refs.~\cite{DeBlas3TeV,IMCC2023}. The markers at $3$, $10$, $14$, and $30$~TeV correspond to $1$, $10$, about $20$, and $90$~ab$^{-1}$. The shaded band indicates the centre-of-mass energy of Layout~B, the collider ring in the 21~km UNK tunnel, with $11$--$16$~T dipoles at $f=0.66$, i.e.\ $14$--$21$~TeV (Table~\ref{tab:unk}).}
\label{fig:lambda3}
\end{figure}

The Higgs potential itself becomes measurable, and here the precision improves strongly with energy (Fig.~\ref{fig:lambda3}). The trilinear self-coupling $\lambda_3$ is probed through double-Higgs production in vector-boson fusion. For $1$~ab$^{-1}$ at $3$~TeV its projected precision is a broad $\sim\!20$--$30\%$, from a non-Gaussian likelihood with two minima, and $\sim\!15\%$ for $2$~ab$^{-1}$~\cite{DeBlas3TeV}. The precision reaches $\approx3.6\%$ for $10$~ab$^{-1}$ at $10$~TeV, $\approx2.5\%$ for about $20$~ab$^{-1}$ at $14$~TeV and $\approx1.2\%$ for $90$~ab$^{-1}$ at $30$~TeV (68\% intervals)~\cite{DeBlas3TeV,IMCC2023}. The first full-simulation analysis at $10$~TeV uses $HH\to b\bar bb\bar b$ alone and includes the machine-induced background. For $10$~ab$^{-1}$ it obtains $0.94<\lambda_3/\lambda_3^{\rm SM}<1.08$ (68\% C.L.), i.e.\ about $7\%$, and $0.97<\lambda_3/\lambda_3^{\rm SM}<1.05$ for two experiments with $10$~ab$^{-1}$ each~\cite{Andreetto2026}. The quartic coupling $\lambda_4$, beyond the reach of any other proposed facility, can be approached through triple-Higgs production. The Standard Model cross section $\sigma(\mu^+\mu^-\to\bar\nu\nu\,HHH)\approx4.2$~ab at $10$~TeV yields about $42$ signal events (before backgrounds and efficiencies) in the $10$~ab$^{-1}$ reference dataset, and $\sim\!84$ in the $20$~ab$^{-1}$ assumed by Ref.~\cite{Chiesa2020quartic}. The yield rises steeply with energy. The quartic term of the Higgs potential could thus be probed, if only loosely, for the first time. Table~\ref{tab:higgs} collects the principal Higgs measurements and their projected precisions.

\begin{table}[!hbt]
\centering
\caption{Representative projected precisions of the main Higgs measurements at a muon collider (statistical, indicative). Here $\delta\lambda_3$ is the precision on the trilinear self-coupling and $\Gamma_H$ the model-independent total width. The Standard Model values are $\Gamma_H=4.1$~MeV at $m_H=125$~GeV~\cite{HXSWGYR4} and $\lambda_3=\lambda_4=m_H^2/2v^2\approx0.13$ for $V\supset\lambda_3 v h^3+\frac{\lambda_4}{4}h^4$ with $v=246$~GeV~\cite{PDG2022}, while $\kappa_\mu=1$ corresponds to $y_\mu=\sqrt{2}\,m_\mu/v\approx6.1\times10^{-4}$. Integrated luminosities are $1$~ab$^{-1}$ at $3$~TeV, $10$~ab$^{-1}$ at $10$~TeV (one experiment), about $20$~ab$^{-1}$ at $14$~TeV and $90$~ab$^{-1}$ at $30$~TeV. $^a$Line-shape scan with $20$~fb$^{-1}$, per decay channel ($b\bar b$ or $WW^*$). $^b$Including HL-LHC.}
\label{tab:higgs}
\begin{tabular}{llcl}\toprule
Observable & Stage & Precision & Reference \\
\midrule
$\sigma(H\to b\bar b)$            & 3~TeV  & $0.75\%$            & \cite{Andreetto2025} \\
$\sigma(H\to b\bar b)$            & 10~TeV & $0.25\%$            & \cite{Andreetto2026} \\
Inclusive Higgs rate             & 10~TeV & $0.75\%$            & \cite{Li2024forward} \\
Total width $\Gamma_H$           & 10~TeV & $(-0.41\%,\,+2.1\%)$ & \cite{Li2024forward} \\
Width $\Gamma_H$ (line shape)$^{\,a}$ & 125~GeV & $\sim\!3\%$ & \cite{deBlasGuLiu2022} \\
$\kappa_\mu$ (global fit)$^{\,b}$ & 10~TeV & $\approx2.4\%$ & \cite{IMCC2025} \\
$\delta\lambda_3$ (trilinear)    & 3~TeV  & $\sim\!20$--$30\%$  & \cite{DeBlas3TeV} \\
$\delta\lambda_3$ (trilinear)    & 10~TeV & $\approx3.6\%$      & \cite{DeBlas3TeV,IMCC2023} \\
$\delta\lambda_3$ (full sim., $b\bar bb\bar b$) & 10~TeV & $\approx7\%$ & \cite{Andreetto2026} \\
$\delta\lambda_3$ (trilinear)    & 14~TeV & $\approx2.5\%$      & \cite{DeBlas3TeV,IMCC2023} \\
$\delta\lambda_3$ (trilinear)    & 30~TeV & $\approx1.2\%$      & \cite{DeBlas3TeV,IMCC2023} \\
$\lambda_4$ (quartic)            & 10~TeV & first access ($\sim\!42$ events in $10$~ab$^{-1}$) & \cite{Chiesa2020quartic} \\
\bottomrule
\end{tabular}
\end{table}

\subsection{Electroweak, top-quark, and SMEFT measurements}

Because VBF and vector-boson scattering dominate, a muon collider is a particularly sensitive probe of the electroweak sector at short distances. The amplitudes for longitudinal-boson scattering grow with energy, and any deviation from the Standard Model cancellation is amplified. Global fits in the Standard Model Effective Field Theory (SMEFT)\footnote{A systematic parametrization of heavy new physics through higher-dimensional operators added to the Standard Model Lagrangian.} translate this growth into sensitivity to new-physics scales of tens of TeV~\cite{Snowmass2022Summary,EWscattering2024}. The top quark is produced abundantly through VBF. A global SMEFT analysis of the top sector improves on the HL-LHC bounds by up to several orders of magnitude for individual four-fermion operators~\cite{TopOperators2025}. The top electroweak dipole moments are constrained at the percent level and below~\cite{TopDipole2024}. To match this experimental accuracy, precision phenomenology at next-to-leading electroweak order is now being developed specifically for multi-TeV muon collisions~\cite{Frixione2025}.

\subsection{Direct and indirect searches for new physics}

Electroweakly charged new states, if they exist, could be pair-produced up to nearly $\sqrt{s}/2$ and singly produced to higher masses through VBF~\cite{ColorfulProd2025}. Studies now cover a broad set of scenarios, among them:
\begin{itemize}
\item Supersymmetric partners, namely sleptons (with privileged access through the muon coupling), charginos, and neutralinos, with thresholds scanned by tuning the beam energy~\cite{FranceschiniGreco2021}.
\item Extended Higgs sectors, such as two-Higgs-doublet models, composite-Higgs resonances, and doubly charged scalars, with a discovery reach extending to several TeV and beam-energy scans of $s$-channel heavy-Higgs resonances~\cite{FranceschiniGreco2021}.
\item Heavy QCD axions, which can be probed through $\mu^+\mu^-\to\mu^+\mu^-a$ over a region of coupling--mass space inaccessible elsewhere~\cite{HeavyAxion2025}.
\item Lepton-flavour-violating effective operators, with effective scales reaching from tens of TeV for dipole and gauge operators up to $\sim\!200$~TeV for four-fermion operators~\cite{LFV2025}, and superheavy Majorana neutrinos that could be produced in VBF~\cite{Majorana2025}.
\item Additional gauge bosons, extra-dimensional Kaluza--Klein gravitons, and anomalous quartic gauge couplings. The last two have been studied explicitly at the $14$~TeV energy relevant to Section~\ref{sec:unk}~\cite{InanKisselevRS2023,Amarkhail2023gamma,Amarkhail2024Zgamma,InanKisselevALP2022}.
\end{itemize}

\subsection{Dark matter and the muon anomaly}

\begin{figure}[!htb]
\centering
\includegraphics[width=0.66\linewidth]{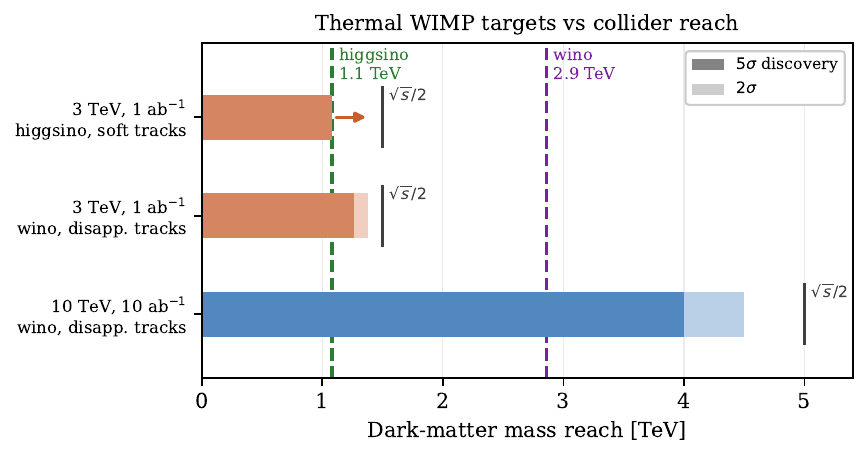}
\caption{Thermal electroweak-WIMP mass targets (higgsino doublet at $1.1$~TeV, wino triplet at $2.9$~TeV~\cite{Bottaro2022wimp,Bottaro2022complex}) compared with the reach of the muon-collider stages. Wino, disappearing tracks at $3$~TeV with $1$~ab$^{-1}$ and at $10$~TeV with $10$~ab$^{-1}$: $5\sigma$ discovery (dark) and $2\sigma$ (light)~\cite{Capdevilla2021tracks}. Higgsino, soft displaced tracks at $3$~TeV with $1$~ab$^{-1}$: a thermal higgsino would be discovered at more than $5\sigma$, and the arrow indicates that the full mass reach is not quoted numerically~\cite{Capdevilla2024soft}. Vertical ticks mark $\sqrt{s}/2$. At $10$~TeV a thermal wino could be discovered already with $\sim\!70$~fb$^{-1}$, less than $1\%$ of the $10$~ab$^{-1}$ dataset.}
\label{fig:dm}
\end{figure}

A particularly clean case is weakly interacting massive particle (WIMP) dark matter, a hypothetical candidate that would have frozen out of thermal equilibrium in the early universe through electroweak-strength interactions. For a candidate that is part of an electroweak multiplet, the measured relic abundance fixes its mass, giving sharp targets: the wino-like triplet at $\approx2.9$~TeV and the higgsino-like doublet at $\approx1.1$~TeV~\cite{Bottaro2022wimp,Bottaro2022complex}. These states are nearly mass-degenerate. The charged partner would therefore travel a few millimetres to centimetres into the detector before decaying into the neutral dark-matter candidate plus an undetectably soft particle. It would leave a short ionization track that stops abruptly, a so-called ``disappearing track''. A $10$~TeV muon collider covers the thermal wino with a large margin. With $10$~ab$^{-1}$ the disappearing-track search reaches wino masses of $4.0$~TeV at $5\sigma$ ($4.5$~TeV at $2\sigma$), close to the kinematic limit $\sqrt{s}/2$. A thermal wino, whose chargino would have $c\tau\approx6$~cm, could be discovered already with $\sim\!70$~fb$^{-1}$, less than $1\%$ of the planned $10$~ab$^{-1}$~\cite{Capdevilla2021tracks}. A thermal higgsino, whose chargino would travel just $c\tau\approx6.6$~mm, is best targeted with soft displaced tracks. This search could discover it at more than $5\sigma$ already at the $3$~TeV stage with $1$~ab$^{-1}$~\cite{Capdevilla2024soft} (Fig.~\ref{fig:dm}). Combined with indirect electroweak-precision measurements~\cite{Franceschini2023,Han2020coupling} and inert-doublet searches~\cite{InertDoublet2024}, a multi-TeV muon collider can close essentially the entire electroweak thermal-WIMP window. Among the other proposals, only FCC-hh is projected to reach both thermal targets with disappearing tracks~\cite{Capdevilla2021tracks}.

Finally, the former $(g-2)_\mu$ tension motivated a related argument. That tension has now been resolved on the theory side. The 2025 White Paper prediction, with the leading-order hadronic vacuum polarization taken from lattice QCD, agrees with the final Fermilab measurement within about one standard deviation~\cite{gm2WP2025,gm2FNAL2025}. An independent hybrid evaluation, lattice QCD with a small long-distance part taken from data where the experiments agree, lies $0.5$ standard deviations from the measured value~\cite{Boccaletti2026HVP}. The tension between some $e^+e^-\to$~hadrons evaluations and the lattice results remains an open question of hadronic physics. The ``no-lose'' argument of Ref.~\cite{Capdevilla2021nolose} therefore holds only conditionally. \emph{If} new physics coupled to the muon is present, a staged muon-collider programme would be guaranteed either to produce the responsible states directly or to reveal their effects in precision channels such as $\mu^+\mu^-\to H\gamma$~\cite{Capdevilla2021nolose,Buttazzo2021}.

\subsection{Frontier opportunities}

Beyond these well-studied cases, the clean environment allows new kinds of measurement. Examples are tests of quantum entanglement and Bell-inequality violation in $\mu^+\mu^-\to ZZ$~\cite{Entanglement2025} and a neutrino programme from the muon decays in the machine (Sections~\ref{sec:ring} and~\ref{sec:applications}). The full-simulation programme is carried out within the Geant4-based detector concepts MAIA (Muon Accelerator Instrumented Apparatus)~\cite{MAIA2025} and MUSIC~\cite{MUSIC2025}. It is now being extended to $\tau$-lepton reconstruction and hence to $H\to\tau^+\tau^-$. Ref.~\cite{Andreetto2025} defers this channel, $H\to c\bar c$ and $H\to Z\gamma$ to forthcoming studies. The $10$~TeV study~\cite{Andreetto2026} also leaves them out, since hadronic $\tau$ reconstruction is not yet implemented and $H\to c\bar c$ is not separated from $H\to b\bar b$.

\section{Technological challenges and proposed solutions}
\label{sec:technology}

\begin{figure}[!htb]
\centering
\includegraphics[width=0.98\linewidth]{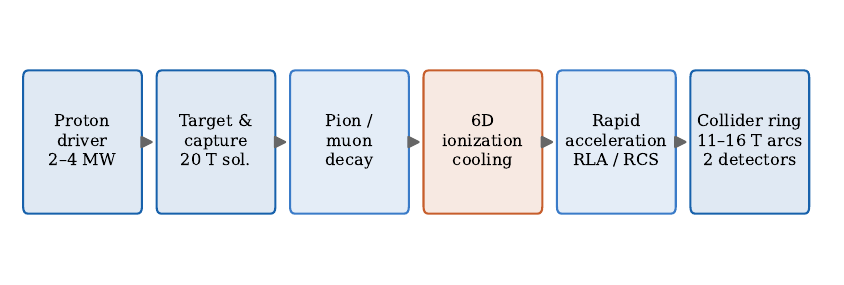}
\caption{Schematic of the muon-collider accelerator complex. A multi-megawatt proton driver produces pions on a target inside a high-field capture solenoid, and the pions decay to muons. The muons are cooled in six dimensions by ionization cooling, rapidly accelerated, and injected into the collider ring. Adapted from the IMCC design~\cite{IMCC2023,MuColParams2025}.}
\label{fig:complex}
\end{figure}

Every subsystem of a muon collider is shaped by the $2.2~\mu\mathrm{s}$ rest lifetime of the muon. At a beam energy of $5$~TeV (the $10$~TeV collider) the Lorentz factor is $\gamma\approx4.7\times10^4$, stretching the laboratory lifetime to about $0.1$~s. In the $11.4$~km reference ring of Table~\ref{tab:params} this corresponds to $\approx2700$ turns. The luminosity decays twice as fast as the beam, so the luminosity lifetime is $\approx1360$ turns~\cite{MuColParams2025,IMCC2025} (Section~\ref{sec:ring}). At the few-hundred-MeV energies of the front end, however, the laboratory lifetime is only a few microseconds. Production and cooling must therefore be completed within microseconds, and the acceleration to $5$~TeV within about ten milliseconds. The full accelerator chain is sketched in Fig.~\ref{fig:complex}, and the design is documented in the IMCC reports~\cite{IMCC2023,IMCCInterim2024,IMCC2025,MuColParams2025,FacilityDesign2022}. For each subsystem the challenge, the proposed solution, and its status and open issues are given in turn. The open issues are collected in Section~\ref{sec:openchallenges}.

\subsection{Muon production and capture}
\label{sec:production}

\emph{Challenge.} Muons must be produced in very large numbers ($\sim\!10^{12}$ per bunch) within a phase space small enough to be cooled. \emph{Proposed solution.} In the IMCC baseline an H$^-$ linac, an accumulator ring, and a compressor ring deliver $5\times10^{14}$ protons per pulse at $5$~GeV and $5$~Hz. This corresponds to $2$~MW, and a higher-power option uses $10$~GeV and $4$~MW. The protons are accumulated in two bunches, shortened by a $90^\circ$ rotation in the longitudinal phase space to an rms length of $2$~ns ($\approx60$~cm), and recombined on the target~\cite{IMCC2025,MuColParams2025}. The baseline target is an $80$~cm isostatic graphite rod, considered adequate up to $2$~MW. Fluidized tungsten and liquid-metal targets are the options for $4$~MW. The graphite absorbs about $110$~kW. This is about four times the $25$~kW deposited in the graphite target of the Long-Baseline Neutrino Facility (LBNF) at a beam power of $1.2$~MW~\cite{IMCC2025,Ammigan2025HPT}. The pions are captured by an HTS solenoid whose field falls from $20$~T at the target to $\approx1.5$~T over the $\approx18$~m capture channel. About $40$~cm of tungsten shields the coils. The captured pions then decay into muons~\cite{IMCC2025}. Target and shielding are being optimized with FLUKA and MARS simulations~\cite{TargetFLUKA2026,FacilityDesign2022}. \emph{Status and open issues.} The front end is specified in the IMCC reports~\cite{IMCCInterim2024,IMCC2025}. Its H$^-$ linac and accumulator ring can be scaled from existing H$^-$ facilities such as the Spallation Neutron Source (SNS), J-PARC, and CERN Linac4~\cite{IMCC2025}. The SNS, operating at $60$~Hz, reached the IMCC baseline power of $2$~MW in 2026~\cite{Hoover2025SNS,ORNL2026SNS}. The two most demanding elements, bunch compression and the target, have not been tested. The compressed bunch lies beyond the space-charge regime of the highest-intensity existing ring, the SNS accumulator ($\sim\!10^{14}$ protons in a bunch $\sim\!150$~m long). At $5$~GeV the simulations still show a space-charge-driven resonance and emittance growth at the end of the rotation, whereas at $10$~GeV the rotation is clean~\cite{IMCC2025,Hoover2025SNS}. On the target side, the IMCC notes that the combination of MW-class power with nanosecond pulses has no precedent. The beam spot, $5$~mm rms, was chosen this large to keep the dynamic stress in the graphite within limits. Even so, the $2$~ns pulse causes an adiabatic temperature rise that may exceed material limits~\cite{IMCC2025}. The MW-class graphite targets of the NuMI (Neutrinos at the Main Injector) beamline and of LBNF work in easier conditions. They receive spills of about $10~\mu$s, on spots of $\approx1.5$~mm rms at NuMI and $1$--$4$~mm rms in the LBNF design, and neither sits inside a superconducting solenoid~\cite{Yonehara2023NuMI,Papadimitriou2016LBNF}. No practical R\&D or beam test of a muon-collider target has been carried out so far, and irradiation tests at the CERN HiRadMat facility are planned~\cite{IMCC2025}. The target technology is therefore not demonstrated.

The two charge signs are not produced symmetrically. Proton--nucleus collisions generate more positive than negative pions, an excess that reflects the positive charge and the valence-quark content of the proton beam and target. For the $5$~GeV IMCC driver about $30\%$ more $\mu^+$ than $\mu^-$ emerge from the target and decay channel, and the captured $\pi^+/\pi^-$ ratio is $\approx1.3$~\cite{IMCC2025,MuColParams2025}. The excess decreases with proton energy. In the same simulations the ratio falls from $\approx1.55$ at $3$~GeV to $\approx1.1$ at $10$~GeV, because the $\pi^+$ yield per unit beam energy decreases while the $\pi^-$ yield stays nearly constant~\cite{IMCC2025,MuColParams2025}. These simulations do not extend beyond $10$~GeV. The total muon yield per unit beam power depends only weakly on the proton energy between $5$ and $15$~GeV~\cite{IMCC2025}. A positive muon beam is therefore easier to produce at high intensity than a negative one. Equal intensities are not required, since the luminosity scales as the product $N_+N_-$ of the bunch populations, but the $\mu^-$ deficit costs luminosity. After longitudinal capture a solenoid chicane separates the two charge signs. The $\mu^+$ and $\mu^-$ bunch trains are then cooled in two parallel channels and recombined before re-acceleration~\cite{IMCC2025}. The split solenoid of the charge separator has no engineering design yet~\cite{IMCC2025}. The charge asymmetry also motivates, as a possible intermediate step, collider modes that use only the more abundant positive muons. A $\mu^+e^-$ and $\mu^+\mu^+$ facility of this kind ($\mu$TRISTAN) has been proposed using ultra-cold $\mu^+$ beams produced by surface-muon cooling~\cite{Hamada2022muTRISTAN}. The same logic could in principle be extended to a $\mu^+p$ mode colliding a positive-muon beam against protons from the driver complex. Such asymmetric modes would not deliver the full $\mu^+\mu^-$ physics programme. They could, however, serve as lower-risk staging options that exercise the muon front end before the symmetric collider is realized. This option is mentioned here only as a direction for discussion, not as part of the baseline design.

\subsection{Six-dimensional ionization cooling}

\emph{Challenge.} The captured muon beam occupies a phase-space volume hundreds of times too large for acceleration. It must be reduced quickly, and ordinary cooling methods (electron, stochastic, laser) are far too slow on the scale of the muon lifetime. \emph{Proposed solution.} In ionization cooling~\cite{AdoBalbekov1971,Skrinsky1981,Neuffer1983} the beam passes through low-$Z$ absorbers (liquid hydrogen or lithium hydride). The ionization energy loss in the absorbers reduces all momentum components, and radio-frequency cavities then restore only the longitudinal one. The beam is cooled in all six phase-space dimensions by iterating absorbers, RF cavities, and strong solenoids. The longitudinal dimension is reached by means of wedge-shaped absorbers that couple energy spread to transverse position. \emph{Status and open issues.} The idea is straightforward, but its practical use is limited by engineering problems. High accelerating gradients must alternate with absorbers inside strong solenoids, and the beam enters the channel with a very large transverse emittance and momentum spread. The MICE experiment observed the ionization-cooling effect. Measuring particle by particle, it found an increase of the phase-space density in the beam core. It also found a reduction of the normalized transverse emittance of muons crossing a single lithium-hydride or liquid-hydrogen absorber, without RF re-acceleration~\cite{MICE2020,MICE2024}. MICE did not, however, demonstrate a practical cooling scheme or address the most challenging engineering issues. The main hardware problem is the RF, since the cavities must sustain high gradients inside strong solenoidal fields. At the MuCool Test Area at Fermilab an $805$~MHz copper cavity that reached $40$~MV/m without an external magnetic field was stable only below $14$~MV/m in $3$~T. The same cavity with beryllium end walls operated stably at $50$~MV/m in $3$~T. The solenoid focuses field-emitted electrons into small spots on the walls, and beryllium, with its low density and stopping power, limits the resulting pulsed heating~\cite{Bowring2020MTA}. High-pressure hydrogen-filled cavities showed no major dependence on fields of up to $3$~T~\cite{Chung2013HPRF}. The IMCC cooling channel needs $20$--$30$~MV/m in $7$--$10$~T, which has not yet been demonstrated. Gas-filled and dielectric-loaded cavities are studied as alternatives to vacuum cavities~\cite{IMCC2025}. An integrated cooling demonstrator with many absorbers, re-acceleration, and a bunched beam is foreseen in the 2030s~\cite{IMCC2025,USWhitePaper2025}.

\subsection{Rapid acceleration}
\label{sec:acceleration}

\emph{Challenge.} The cooled muons must be accelerated from $255$~MeV to $5$~TeV before a substantial fraction decays. \emph{Proposed solution.} A cascade of a single-pass linac and two recirculating linear accelerators (RLAs) with $352$~MHz superconducting cavities brings the muons to $63$~GeV. Four rapid-cycling synchrotrons (RCS) follow. The first has normal-conducting magnets only. The other three are hybrid, with fixed superconducting dipoles interleaved with normal-conducting dipoles that swing from $-1.8$ to $+1.8$~T~\cite{IMCC2025}. The four rings take $0.34$, $1.10$, $2.37$, and $6.37$~ms, about $10$~ms in total, and the ramp rates fall from $\approx4200$~T/s in the first ring to $\approx565$~T/s in the last~\cite{IMCC2025,MuColParams2025}. The acceleration chain contains the longest tunnels of the complex and by far the largest RF voltage of its accelerating stages. The four rings have circumferences of $6.0$, $6.0$, $10.7$, and $35$~km, and the last one gains $63.6$~GeV per turn~\cite{IMCC2025,MuColParams2025}. The first two rings share one tunnel~\cite{MuColParams2025}. The four synchrotrons together need $\approx133$~GV of RF voltage in $\approx5.2$~km of active cavity length. The last ring alone has $72.7$~GV in $2.8$~km, whereas the second RLA has $\approx15$~GV~\cite[Tables~I.6 and~I.8]{MuColParams2025}. Most of the arc length of the last ring is taken by normal-conducting dipoles that reach only $1.8$~T. Its average bending field at extraction is thus only about $3$~T (an estimate from the IMCC ring parameters), so the ring is about three times longer than the $11.4$~km collider ring. \emph{Status and open issues.} The RLAs follow the multi-pass scheme of CEBAF ($12$~GeV in $5.5$~passes with $1497$~MHz cavities~\cite{Adderley2024CEBAF}). The large phase space of the muon beam, however, requires low-frequency cavities, large apertures, and tight focusing. So far only the second RLA has a lattice design~\cite{IMCC2025,MuColParams2025}. In hardware, the rapid-cycling synchrotrons are the least mature stage of the acceleration chain. Operating proton rapid-cycling synchrotrons ramp their dipoles sinusoidally. Their peak rates are about $30$~T/s at the Fermilab Booster ($15$~Hz), $65$--$70$~T/s at the J-PARC RCS ($25$~Hz), and about $80$~T/s at ISIS ($50$~Hz)~\cite{FNALBoosterRookie1998,Yoshii2008JPARCRF,MagnetRD2025,ISISGuide2025}. The muon rings need nearly linear ramps at mean rates of $565$--$4200$~T/s, some $7$ to $50$ times the peak rate of ISIS~\cite{IMCC2025,MuColParams2025}. A fast-cycling HTS test magnet has reached $\approx300$~T/s over a $0.5$~T field span~\cite{Piekarz2022HTS}, but no magnet system has been demonstrated at this combination of field and ramp rate~\cite{Bottura2025MagnetRD}. A string of $1.8$~T magnets with energy recovery, due in 2032, is planned as the first system test within the proposed ten-year IMCC magnet programme~\cite{MagnetRD2025}. This review regards its milestones as ambitious.

\subsection{High-field superconducting magnets}
\label{subsec:magnets}

\emph{Challenge.} Three distinct magnet families are needed at fields beyond present accelerator practice. The first is a $20$~T capture solenoid with a free bore of at least $1.4$~m. The final cooling cells need solenoids of at least $40$~T in a $50$~mm bore, and the collider ring needs arc dipoles of $14$--$16$~T~\cite{IMCC2025,Bottura2025MagnetRD,MuColParams2025}. \emph{Proposed solution.} High-temperature superconductors, in particular rare-earth--barium--copper oxide (REBCO), are the enabling material. The IMCC magnet programme proposes a ten-year development plan for all three families. Its resource needs are estimated at about $83$~million Swiss francs in materials and $\sim\!400$ full-time-equivalent person-years of effort~\cite{MagnetRD2025}. \emph{Status and open issues.} Small-bore d.c. solenoids have reached fields of the required magnitude. The $32$~T all-superconducting user magnet of the US National High Magnetic Field Laboratory has a $34$~mm cold bore, and its HTS insert is under repair after damage found in 2023. A field of $45.5$~T has been reached in a solenoid with a $14$~mm-bore REBCO insert in a $31.1$~T resistive background~\cite{Hahn2019,Bottura2025MagnetRD}. No laboratory magnet yet combines the $50$~mm bore and $\sim\!40$~T needed for final cooling, which the IMCC plans to demonstrate by 2034. HTS magnets for the muon collider are at technology-readiness level (TRL) $3$--$4$, versus a target of $6$~\cite{MagnetRD2025}. Quench protection is a major unresolved issue for these magnets. Their stored energies reach several to tens of megajoules per magnet at energy densities up to $300$~J/cm$^3$. Insulated HTS coils at such densities are very difficult to protect against localized quenches, and the non-insulated windings proposed as a remedy have still to be demonstrated in accelerator magnets~\cite{MagnetRD2025,Bottura2025MagnetRD,IMCC2025}. All-HTS accelerator-magnet designs are under development within the IMCC and the US Magnet Development Program~\cite{MagnetRD2025,USMDP2025}, with a wide-aperture $14$~T HTS dipole demonstrator foreseen by 2045~\cite{MagnetRD2025}. Turning laboratory fields into accelerator-quality, mass-producible magnets is the principal hardware R\&D item of the project. The schedule and resource figures quoted here are IMCC estimates.

\subsection{The collider ring: turns, luminosity, and neutrino flux}
\label{sec:ring}

The energy reach of a circular collider is set by how strongly its magnets bend the beam. The magnetic-rigidity relation~\cite{PDG2022},
\begin{equation}
p\,[\mathrm{GeV}/c] = 0.2998\;B\,[\mathrm{T}]\;\rho\,[\mathrm{m}],
\label{eq:rigidity}
\end{equation}
gives the beam momentum $p$ in terms of the dipole field $B$ and the bending radius $\rho$. With two beams of momentum $p$ colliding head-on, $\sqrt{s} = 2p$. The same relation fixes the number of turns that a stored muon beam survives. Because the stored muons decay, a fill is useful only for the number of revolutions the beam makes within its dilated lifetime,
\begin{equation}
n_{\rm turns} \simeq \frac{\gamma\,c\tau_\mu}{C},
\label{eq:nturns}
\end{equation}
where $\gamma=E_\mu/m_\mu c^2$ is the beam Lorentz factor, $c\tau_\mu = 659$~m, and $C$ is the circumference. Applied around the whole ring, Eq.~\eqref{eq:rigidity} gives $E_\mu\,[\mathrm{GeV}] = 0.2998\,\bar B\,(C/2\pi)$, where $\bar B$ is the circumference-averaged bending field. Substituting this into Eq.~\eqref{eq:nturns}, the circumference cancels:
\begin{equation}
n_{\rm turns} \simeq \frac{0.2998\,\bar B\,c\tau_\mu}{2\pi\,m_\mu c^2\,[\mathrm{GeV}]} \approx 297\,\bar B\,[\mathrm{T}].
\label{eq:nturnsB}
\end{equation}
The number of stored turns therefore depends only on the average field. At fixed $\bar B$ it does not depend on the beam energy, because the circumference grows in proportion to the energy. The luminosity is proportional to the product of the two decaying bunch populations and so decays twice as fast as the beams. The luminosity lifetime is half of Eq.~\eqref{eq:nturnsB}, $\approx149\,\bar B$[T] turns. With $14$~T arc dipoles, as in the $11.4$~km reference ring at $10$~TeV, $\bar B\approx9.2$~T and the luminosity lifetime is $\approx1360$ turns~\cite{MuColParams2025,IMCC2025}. The IMCC parameter tables follow this rule. The $4.5$~km ring at $3$~TeV and a $15$~km variant of the $10$~TeV ring both have $\bar B\approx7$~T and the same luminosity lifetime of $\approx1040$ turns~\cite{IMCC2025}.

The time-averaged luminosity is obtained by summing the bunch crossings of one fill and multiplying by the repetition rate $f_r$ of the complex. Each crossing is weighted by the surviving populations $N^2e^{-2j/n_{\rm turns}}$ of the two decaying beams~\cite[Sec.~3.1, Eqs.~(8)--(10)]{IMCC2023}:
\begin{equation}
\bar{\mathcal L} \simeq f_r\,n_b\,\frac{N^2}{4\pi\sigma^{*2}}\sum_{j\ge0}e^{-2j/n_{\rm turns}}
\simeq f_r\,n_b\,\frac{N^2}{8\pi\sigma^{*2}}\,n_{\rm turns},
\label{eq:Lavg}
\end{equation}
where $n_b$ is the number of bunches per beam, $N$ the number of muons per bunch at injection, and $\sigma^*$ the rms beam size at the interaction point. The revolution frequency $f_{\rm rev}=c/C$ does not enter as a separate factor, because it is already contained in $n_{\rm turns}=f_{\rm rev}\,\gamma\tau_\mu$. A smaller ring has a proportionally higher peak luminosity, $n_bN^2f_{\rm rev}/4\pi\sigma^{*2}$, but its fill is proportionally shorter. At a fixed repetition rate only the number of crossings per fill counts. The MuCol luminosity formula has the same structure, with $f_{\rm rev}$ multiplied by a decay-averaging factor $\propto f_r\gamma\tau_\mu$~\cite[Eqs.~(A.5)--(A.8)]{MuColParams2025}. Together with Eq.~\eqref{eq:nturnsB}, Eq.~\eqref{eq:Lavg} gives $\bar{\mathcal L}\propto\bar B$ at fixed $f_r$, $n_b$, $N$ and $\sigma^*$, the form given explicitly in Ref.~\cite{IMCC2023}. At \emph{fixed energy} the beam size $\sigma^*$, set by the emittance, $\beta^*$ and $\gamma$, is unchanged, while the rigidity relation gives $\bar B\propto 1/C$. Hence
\begin{equation}
\bar{\mathcal L}\big|_{\text{fixed }\sqrt{s}} \;\propto\; \bar B \;\propto\; \frac{1}{C}.
\label{eq:Lscaling}
\end{equation}
The IMCC parameter table follows this scaling. At $\sqrt{s}=10$~TeV and identical beam parameters, the estimated luminosity is $21\times10^{34}$~cm$^{-2}$\,s$^{-1}$ for a $10$~km ring and $14\times10^{34}$~cm$^{-2}$\,s$^{-1}$ for a $15$~km ring~\cite{IMCC2025}. The MuCol consolidated-parameters report draws the design consequence: maximizing the luminosity requires the shortest circumference compatible with the feasibility of the magnets~\cite[Sec.~6]{MuColParams2025}.

These estimates include only the muon decay. Other mechanisms of beam loss and luminosity degradation are not included and are still under study~\cite{IMCC2025}. They include transverse instabilities driven by the resistive-wall impedance and the head-on beam--beam instability, which the IMCC mitigates in simulation with a non-zero chromaticity and a transverse damper. They also include the insufficient momentum acceptance of the present interaction-region lattices (Section~\ref{sec:openchallenges}), halo losses on the aperture, and machine imperfections, which have not yet been studied~\cite{IMCC2025,MuColParams2025}. Keeping the bunches short for about $1000$ turns is a further challenge~\cite{IMCC2025}.

The decays also emit neutrinos at angles of order $1/\gamma$ to the beam direction, only $\approx2\times10^{-5}$~rad at $5$~TeV. In the arcs the emission direction turns with the beam, so the flux forms a thin disc in the plane of the ring. The neutrinos cross the Earth almost unattenuated. The radiation dose arises far from the ring, where this disc emerges at the surface and the neutrinos induce showers in the ground and in buildings~\cite{IMCC2025}. For a ring at depth $h$, the decay-plane neutrinos emitted tangent to the Earth's surface emerge at a distance $d\simeq\sqrt{2hR_\oplus}$, where $R_\oplus\approx6371$~km is the Earth's radius. Each straight section emits a concentrated flux in one fixed direction. The IMCC design minimizes the length of the straight sections outside the one that houses the interaction points. The surfacing areas of the experimental straight sections are to be owned and fenced by the operating laboratory~\cite{IMCC2025}. The flux from the arcs is diluted by a periodic vertical deformation of the ring, which varies the vertical slope of the orbit within $\pm1$~mrad (Section~\ref{sec:openchallenges}). This spreads the flux over a larger surface area and reduces the peak dose by a factor of $80$--$90$~\cite{IMCC2023,IMCCInterim2024,IMCC2025}. A ring that is not deep underground needs additional horizontal closed-orbit deformations~\cite{IMCC2025}. The flux becomes more forward-peaked and each neutrino more energetic as the beam energy grows. Per muon decay, the off-site dose from a bending section with the vertical deformation rises approximately as the square of the beam energy~\cite{MuColParams2025}. Without the deformation the flux keeps its natural vertical width of order $1/\gamma$. The dose from such an idealized ring then rises as $E_\mu^3$ at a fixed muon rate, a conservative estimate that takes the neutrino cross section as proportional to the energy~\cite{NeufferShiltsev2018}. In a straight section the flux is narrow in both transverse directions. The peak dose per decay from such a section rises faster still, about as $E_\mu^{3.6}$ between $1.5$ and $5$~TeV (an estimate from the dose kernels of Ref.~\cite[Tables~13.1 and~13.2]{MuColParams2025}). The neutrino flux therefore becomes a leading siting constraint at and above $10$~TeV. The IMCC objective is an effective dose to the public not exceeding $\approx10~\mu$Sv per year, comparable to the impact of the LHC~\cite{IMCC2025,MuColParams2025}.

\subsection{Machine--detector interface and beam-induced background}

\emph{Challenge.} The muons stored in the collider ring decay continuously, with about $6\times10^4$ decays per metre for every bunch passage at the 10~TeV design (some $2\times10^9$ decays per metre per second). The decay electrons are nearly collinear with the beam and shower in the beam pipe and magnets. The showers spray the detector with secondary electrons, photons, and neutrons, the beam-induced background. \emph{Proposed solution.} Two cone-shaped tungsten-alloy nozzles clad with borated polyethylene, with a $10^\circ$ opening angle, are placed on either side of the interaction point. They absorb most of the showers, reduce the flux into the detector by several orders of magnitude, and define the inner envelope and angular acceptance of the detector~\cite{Bartosik2022,IMCC2025}. Precise (sub-nanosecond) timing rejects the out-of-time background, and high-granularity detectors add directional information~\cite{Bartosik2022,BIBreduction2022,Casarsa2024}. A further background source becomes important at $\sqrt{s}\gtrsim10$~TeV. Incoherent $e^+e^-$ pair production by the colliding beams at the interaction point has been identified and characterized~\cite{Casarsa2024}. \emph{Status and open issues.} Full-simulation studies show that the physics performance is preserved in the presence of the beam-induced background at $3$~TeV~\cite{Andreetto2025,Bartosik2022}. At $10$~TeV, with the incoherent pairs also overlaid, sub-percent statistical precision on the leading Higgs channels is retained~\cite{Andreetto2026}. These results rest on simulation. The neutrino flux, the other consequence of the decays, is a major input to siting (Sections~\ref{sec:ring} and~\ref{sec:unk}).

\subsection{Detector concepts}

Two detector concepts, MAIA and MUSIC (Section~\ref{sec:physics}), have been developed using full Geant4 simulation. The MAIA (Muon Accelerator Instrumented Apparatus) concept has an all-silicon tracker in a $5$~T solenoid, silicon--tungsten and iron--scintillator calorimetry, and an external muon spectrometer~\cite{MAIA2025}. MUSIC is an INFN-led alternative with a detailed analysis of the material budget and of the detector occupancy from the beam-induced background~\cite{MUSIC2025}. Both achieve object-reconstruction efficiencies above $95\%$ in the presence of this background, and both incorporate the forward-muon coverage required for the Higgs-width measurement~\cite{ForwardMuon2024}.

Table~\ref{tab:tech} summarizes the principal challenges, the solution proposed for each, and its present status. Most of the solutions exist as designs studied in detailed simulation. None of the critical systems has yet been demonstrated at the required parameters, and the open issues are collected in Section~\ref{sec:openchallenges}.

\begin{table}[!hbt]
\centering
\caption{Principal technological challenges of a muon collider, the proposed solution, and its status and open issues.}
\label{tab:tech}
\begin{tabular}{>{\raggedright\arraybackslash}p{3.0cm}>{\raggedright\arraybackslash}p{4.6cm}>{\raggedright\arraybackslash}p{7.4cm}}\toprule
Challenge & Proposed solution & Status and open issues \\
\midrule
Muon lifetime ($2.2~\mu$s) & Relativistic dilation, chain completed in ms & Not demonstrated as an integrated chain~\cite{IMCC2023} \\ \addlinespace[5pt]
Production of ${\sim}10^{12}\,\mu$/bunch & MW proton driver, target, $20$~T capture solenoid & Not demonstrated. Bunch compression and target untested, existing MW targets use $\sim\!10~\mu$s spills without a solenoid~\cite{IMCC2025,TargetFLUKA2026,Yonehara2023NuMI} \\ \addlinespace[5pt]
Six-dimensional (6D) phase-space reduction & Ionization cooling & Not demonstrated. MICE: transverse emittance reduction in one absorber, no re-acceleration~\cite{MICE2020,MICE2024}. RF in $7$--$10$~T and 6D demonstrator pending~\cite{IMCC2025} \\ \addlinespace[5pt]
Rapid acceleration & Linacs, two RLAs, and rapid-cycling synchrotrons & Not demonstrated. Lattice concepts only, choice of kT/s hybrid magnets open~\cite{IMCC2025,MagnetRD2025} \\ \addlinespace[5pt]
High-field magnets & HTS (REBCO): $\sim\!40$~T solenoids, $14$--$16$~T dipoles & Not demonstrated. TRL $3$--$4$, $>30$~T only in small-bore solenoids, quench protection unresolved~\cite{MagnetRD2025,Hahn2019} \\ \addlinespace[5pt]
Collider ring and interaction region & $\beta^*=1.5$~mm optics and tungsten liner & Not demonstrated. Momentum acceptance insufficient, $\approx450$--$500$~W/m of decay heat~\cite{IMCC2025,MuColParams2025} \\ \addlinespace[5pt]
Beam-induced background & Tungsten nozzles, ns timing, granular detectors & Not demonstrated. Studied in full simulation, incl.\ $10$~TeV Higgs analyses~\cite{Bartosik2022,Casarsa2024,Andreetto2026} \\ \addlinespace[5pt]
Neutrino radiation (grows with $\sqrt{s}$) & Deep siting, periodic vertical ring deformation & Not demonstrated. Concept defined on paper, compatibility with the chromatic correction not shown, harder above $10$~TeV~\cite{IMCC2025} \\ \addlinespace[5pt]
Detection in the background & MAIA, MUSIC concepts & Not demonstrated. $>95\%$ efficiency in simulation~\cite{MAIA2025,MUSIC2025} \\ \addlinespace[5pt]
Design luminosity & All specifications met at once & Not demonstrated. Design target (cf.\ SLC, Section~\ref{sec:openchallenges})~\cite{IMCC2025,Phinney2000SLC} \\
\bottomrule
\end{tabular}
\end{table}

\subsection{Open accelerator challenges}
\label{sec:openchallenges}

The design luminosity requires every stage of the chain to meet its specification at the same time. In the assessment of this review, the following items remain open.

\emph{Proton bunch compression.} Compressing $2.5\times10^{14}$ protons per bunch to $2$~ns rms implies a peak longitudinal charge density several hundred times that of the SNS ring (an estimate from the numbers of Section~\ref{sec:production}). There $\sim\!10^{14}$ protons occupy $\approx550$~ns ($\approx150$~m). The two compressed bunches must then be recombined on the target. A compression test in the SNS ring has been proposed~\cite{IMCC2025,Hoover2025SNS}.

\emph{MW target inside a superconducting solenoid.} A MW-class target hit by nanosecond pulses inside a $20$~T superconducting solenoid has no operating precedent. No practical R\&D or beam test of the muon-collider target has been carried out so far~\cite{IMCC2025}.

\emph{RF in strong solenoidal fields.} Cooling needs $20$--$30$~MV/m in $7$--$10$~T, whereas stable high-gradient operation has been shown only up to $3$~T~\cite{Bowring2020MTA,IMCC2025}.

\emph{Fast-ramping magnets.} The first synchrotron ramps at $\approx4200$~T/s, versus peak rates of $65$--$80$~T/s in operating rapid-cycling synchrotrons such as J-PARC and ISIS (Section~\ref{sec:acceleration})~\cite{IMCC2025,MagnetRD2025,ISISGuide2025}.

\emph{Interaction-region optics.} At $10$~TeV the design requires $\beta^*=1.5$~mm in both planes. This is a hundred times smaller than the $0.15$~m of the HL-LHC baseline, and the nominal LHC value is $0.55$~m~\cite{IMCC2025,Apollinari2015HLLHC}. Values of this order are used only in the vertical plane of the flat beams of the SuperKEKB $e^+e^-$ collider. It has operated routinely at $\beta^*_y\approx1$~mm since 2019, whereas its design value is $\approx0.3$~mm~\cite{Zhou2024SuperKEKB,Zimmermann2025SuperKEKB}. The muon design also has an rms momentum spread of $10^{-3}$, about nine times that of the LHC~\cite{IMCC2025,Apollinari2015HLLHC}. The lattices developed so far for this $\beta^*$ lack sufficient momentum acceptance even for a perfect machine. A $\beta^*$ twice as large would allow a lattice that works for a perfect machine. The IMCC notes, however, that its assumptions on the feasible fields and gradients of the large-aperture magnets have been somewhat optimistic~\cite{IMCC2025}. By extrapolation from other projects with small $\beta^*$ and local chromatic compensation, the IMCC also expects very tight tolerances~\cite{IMCC2025}. These requirements can already be stated, and they lie well beyond those of operating hadron colliders.

\emph{Decay heat in the collider magnets.} At $10$~TeV the decay electrons carry $\approx450$--$500$~W/m, depending on the ring circumference~\cite{IMCC2025,MuColParams2025}. The load per metre scales as $E_\mu/C$, so it depends on the ring layout (Section~\ref{subsec:unkreach}). The cold mass can take $O(5)$~W/m at $4.2$~K or $O(10)$~W/m at $20$~K, so a tungsten liner of $3$--$4$~cm must absorb all but $4$--$8$~W/m~\cite{IMCC2025}. The liner sets the arc-dipole aperture, at least $158$~mm for a $4.5$~K cold mass and $138$~mm at $20$~K. In these apertures the IMCC targets $11$~T with Nb$_3$Sn and $14$~T with REBCO~\cite{IMCC2025}.

\emph{Quench protection of HTS magnets.} Protecting high-energy-density HTS coils against localized quenches is an unresolved issue for both the cooling solenoids and the collider dipoles (Section~\ref{subsec:magnets})~\cite{MagnetRD2025,IMCC2025}.

\emph{Ring deformation for neutrino-dose mitigation.} The magnets of the arcs are moved so that the orbit follows opposing parabolas between the interaction points, with its vertical slope varying within $\pm1$~mrad. For a $100$~m period this means about $\pm25$~mm of vertical movement. About $200$ deformation patterns are applied, each for about $12$~h, and every step interrupts the beam circulation. Orbit variation inside the beam pipe would suffice at $1.5$~TeV beam energy but not at $5$~TeV. The deformation also reaches the chromatic-correction sections. There the strong sextupoles turn the induced vertical dispersion into momentum-dependent coupling. The IMCC rates this study as its highest priority. It also notes that the resulting relative movements of adjacent magnets and of their interconnections have never been seen in an accelerator~\cite{IMCC2025}.

\emph{Delivering the luminosity.} The luminosities of Table~\ref{tab:params} assume that all these specifications are reached~\cite{IMCC2025,MuColParams2025}. Design luminosities of new colliders have not always been reached. The Stanford Linear Collider (SLC) reached a peak luminosity of $3\times10^{30}$~cm$^{-2}$s$^{-1}$, about half of its design value, only in its final 1997--98 run, after about ten years of operation~\cite{Phinney2000SLC,FCCeeQA2019}. The Electron--Ion Collider (EIC) project has deferred the strong hadron cooling on which its $10^{34}$~cm$^{-2}$s$^{-1}$ goal relied~\cite{Willeke2021EICCDR,Wang2026EICSHC}. The luminosities used in Sections~\ref{sec:physics} and~\ref{sec:unk} are design targets in this sense.

Most of the above-mentioned challenges have a planned test. The tests are bunch compression in the SNS ring, target irradiation at HiRadMat, an RF test stand with a $10$--$15$~T solenoid, the RCS magnet string, the $40$~T solenoid, and the integrated cooling demonstrator~\cite{IMCC2025,MagnetRD2025}. No hardware test is planned for the interaction-region optics, and the luminosity itself can be established only in operation.

\subsection{Roadmap, cost, and power}

The IMCC plan, which assumes that these R\&D items are resolved on schedule, foresees a cooling demonstrator in the 2030s and a conceptual design report later in the decade. A first collider stage would operate around mid-century, consistent with the P5 and National Academies recommendations~\cite{P5Report2023,NASEM2025,USInput2026}. The 2026 update of the European Strategy for Particle Physics was adopted by the CERN Council in May 2026. It recommended the $e^+e^-$ Future Circular Collider (FCC-ee) as the preferred option for the next flagship at CERN. It identified the muon collider as an alternative path towards high-energy lepton collisions whose key technologies, above all six-dimensional cooling, still require demonstration, and recommended that the corresponding R\&D be pursued~\cite{ESPP2026,ComparativeEval2025}. The projected site power, $\approx230$~MW at $3$~TeV and $\approx300$~MW at $10$~TeV (Table~\ref{tab:params}), is comparable to that of the large $e^+e^-$ factories, whose energy reach is far lower. It is roughly half that of a $100$~TeV proton collider of comparable physics reach. The IMCC cost estimate, made for the option at CERN, is broken down by technical category rather than by subsystem (Section~\ref{subsec:unkinfra})~\cite{IMCC2025}. Energy efficiency and a relatively compact size remain among the collider's strongest practical arguments~\cite{ITF2022,Stratakis2025}.

\section{A muon collider and the 21~km UNK tunnel}
\label{sec:unk}

\subsection{The UNK tunnel}
\label{subsec:unktunnel}

In the 1980s and early 1990s IHEP excavated the tunnel of the Accelerating and Storage Complex (UNK) near the town of Protvino. It is a ring of circumference $C = 20{,}772$~m ($\approx21$~km) and $5.1$~m in diameter~\cite{GurovUNK1995}. The site, chosen in 1982 after geological studies, places it in stable limestone $20$--$60$~m below the surface~\cite{Balbekov1983}. The tunnel was designed to house a proton accelerator reaching the TeV scale, but the project was halted in the 1990s. The ring tunnel had been completely excavated by 1995~\cite{GurovUNK1995} but was never equipped. The 1990--91 design has six matched straight sections. Four of them, $490$~m long, were reserved for experiments, and two of $800$~m for injection, beam abort, RF, and extraction~\cite{Yarba1991UNK}. Together they take $3.56$~km, about $17\%$ of the circumference. For a muon collider the dominant cost and technical risk lie in the accelerator systems rather than in the tunnel. These are the proton driver and target, the cooling channel and its RF, the rapid-cycling synchrotrons, and the high-field magnets. A tunnel of this size is nevertheless a substantial civil-engineering cost and a long-lead-time item. At Protvino it already exists, and Section~\ref{subsec:layouts} discusses which part of the complex it could house.

The idea of placing a muon collider in an existing tunnel is not new. Neuffer and Shiltsev analysed a pulsed $14$~TeV muon collider in the $27$~km LHC tunnel and found it feasible in principle. In their scheme the collider ring and a hybrid rapid-cycling synchrotron share that tunnel. The synchrotron reaches $7$~TeV with $16$~T fixed and up to $3.8$~T pulsed dipoles at a packing factor of $0.85$, accepting $50\%$ muon survival per stage~\cite{NeufferShiltsev2018}. The IMCC option at CERN reuses the existing tunnels differently. The SPS and LHC tunnels would house the rapid-cycling synchrotrons, and a new $10$~km collider ring would reach $\sqrt{s}=7.6$~TeV~\cite{IMCC2025}. Layouts~A and~B below keep the collider ring and the last synchrotron in separate tunnels, as the IMCC option does. Layout~C places both in the UNK tunnel, as Neuffer and Shiltsev did for the LHC.

\subsection{The acceleration chain and three layouts}
\label{subsec:layouts}

A collider ring is fed by an acceleration chain that must reach the full beam energy. The last synchrotron of the chain is longer than the collider ring, because it accelerates to the top energy with a lower average field. In the IMCC greenfield design it is $35$~km in circumference, three times the $11.4$~km collider ring, and its average bending field at extraction is about $3$~T (Section~\ref{sec:acceleration}). A simple estimate shows the scale of its RF. An energy gain of order $100$~GeV per turn (the last IMCC ring gains $63.6$~GeV~\cite{IMCC2025}) at a real-estate gradient of $20$~MV/m needs some $5$~km of cavities. This agrees in order of magnitude with the IMCC chain, whose four synchrotrons need $\approx133$~GV in $\approx5.2$~km of active cavity length (Section~\ref{sec:acceleration}). A $10$~TeV greenfield facility needs about $70$~km of tunnels in total. The CERN option at $7.6$~TeV needs only $15$~km of new tunnels, because the SPS and LHC tunnels house the synchrotrons~\cite{IMCC2025}. The circumference of a synchrotron that reaches the beam energy $E_\mu$ with an average field $\bar B$ follows from Eq.~\eqref{eq:rigidity},
\begin{equation}
C_{\rm RCS}\,[\mathrm{m}] = \frac{2\pi\,E_\mu\,[\mathrm{GeV}]}{0.2998\,\bar B\,[\mathrm{T}]},
\label{eq:crcs}
\end{equation}
and the same relation fixes the collider ring for the average field of its arcs. Three layouts follow for the UNK tunnel.

\emph{Layout A: the UNK tunnel as the tunnel of the last synchrotron.} This layout follows the IMCC option at CERN, where the last synchrotron, in the $26.7$~km LHC tunnel, reaches $3.8$~TeV per beam~\cite{IMCC2025}. The average field of that ring at extraction is about $3$~T, as in the greenfield design (an estimate from the IMCC ring parameters). With the same field in the UNK tunnel, Eq.~\eqref{eq:crcs} gives $E_\mu\approx3.0$~TeV, i.e.\ $\sqrt{s}\approx6$~TeV. This is about $20\%$ below the CERN option, in proportion to the shorter tunnel. The collider ring for this energy would be about $7$~km long with $14$~T dipoles ($\bar B\approx9.2$~T) and about $9$~km with $11$~T Nb$_3$Sn dipoles ($\bar B\approx7.3$~T). This new tunnel would be a third to two fifths of the UNK circumference. A higher energy would require stronger fast-ramping magnets. The IMCC mentions fast-ramping HTS dipoles in the last synchrotron as a way to raise the energy of the CERN option~\cite{IMCC2025}, but such magnets have not been built. In the CERN option only the first synchrotron occupies the SPS ($6.9$~km), while the second and the hybrid third synchrotron share the LHC tunnel~\cite{IMCC2025}. The analogue of Layout~A would therefore place two synchrotrons in the UNK tunnel and need a separate ring of some $6$--$7$~km for the first one. No study has yet addressed this arrangement for the UNK tunnel.

\emph{Layout B: the UNK tunnel as the collider ring.} This layout gives the highest energy for a given magnet technology, $\sqrt{s}\approx14$--$21$~TeV for arc dipoles of $11$--$16$~T (Section~\ref{subsec:energyB}). The last synchrotron must then reach $7.2$--$10.5$~TeV per beam. At $\bar B\approx3$~T, Eq.~\eqref{eq:crcs} gives circumferences of about $50$, $64$, and $73$~km for $\sqrt{s}=14.4$, $18.3$, and $20.9$~TeV. This is a new tunnel $2.4$ to $3.5$ times longer than the UNK ring itself. Compared with the greenfield design, Layout~B replaces the $35$~km synchrotron by this longer ring and the collider ring of about $11$~km by the UNK tunnel. It would thus need more new tunnel in total, about $74$--$97$~km against $\approx70$~km (an estimate from these lengths). At fixed muon survival and a fixed ratio of extraction to injection energy, the energy gain per turn scales with the ring circumference~\cite{NeufferShiltsev2018}. The RF of this ring therefore grows in proportion to its top energy. Scaled from the last IMCC ring, it would need about $90$--$130$~GeV per turn, or roughly $105$--$150$~GV of RF in that ring alone (an estimate by this scaling)~\cite[Table~I.8]{MuColParams2025}. A synchrotron in the UNK tunnel itself would need an average field of $7.3$--$10.6$~T. This is well above the $\approx5.5$~T of the Neuffer--Shiltsev synchrotron (an estimate from Eq.~\eqref{eq:crcs}). The six UNK straight sections then belong to the collider ring, where they matter most for the neutrino dose (Section~\ref{subsec:unkinfra}).

\emph{Layout C: the collider ring and the last synchrotron in the UNK tunnel.} This layout follows the scheme of Neuffer and Shiltsev for the LHC tunnel (Section~\ref{subsec:unktunnel})~\cite{NeufferShiltsev2018}. Rings sharing one tunnel have precedents. In the IMCC option at CERN the second and third synchrotrons share the LHC tunnel~\cite{IMCC2025}. The UNK design placed two accelerators in the same $5.1$~m tunnel, the conventional ring UNK-1 above the superconducting ring UNK-2. It left space for a second superconducting ring, UNK-3, intended for $3\times3$~TeV proton--proton collisions~\cite{GurovUNK1995,Yarba1991UNK}. The LHC arc tunnel is narrower, with a finished internal diameter of $3.7$~m~\cite{EvansBryant2008}. In a hybrid synchrotron, fixed superconducting dipoles of field $B_{\rm SC}$ alternate with pulsed dipoles that swing from $-B_{\rm p}$ to $+B_{\rm p}$ during the ramp. If the dipoles fill a fraction $\eta$ of the circumference $C$, the bending at injection and at extraction fixes the extraction energy~\cite{NeufferShiltsev2018},
\begin{equation}
E_{\max}\,[\mathrm{GeV}] = \frac{0.2998\,\eta\,C\,[\mathrm{m}]\,B_{\rm SC}B_{\rm p}/\pi + (B_{\rm SC}-B_{\rm p})\,E_{\rm inj}\,[\mathrm{GeV}]}{B_{\rm SC}+B_{\rm p}},
\label{eq:hybridrcs}
\end{equation}
with the fields in tesla and $E_{\rm inj}$ the injection energy. With $\eta=0.7$, the upper limit that the IMCC adopts for the LHC tunnel, Eq.~\eqref{eq:hybridrcs} reproduces the IMCC extraction energies in that tunnel. For injection at $1.6$~TeV it gives $3.8$~TeV with $10$~T fixed and $\pm1.8$~T normal-conducting dipoles, and $5.0$~TeV with $\pm3$~T pulsed HTS dipoles, as in Ref.~\cite[Tables~14.3.1 and~14.3.3]{IMCC2025}. For the UNK tunnel, with injection at $1.5$~TeV as into the last IMCC synchrotron, it gives three points. The IMCC baseline magnets give $E_\mu\approx3.2$~TeV, i.e.\ $\sqrt{s}\approx6$~TeV. This is slightly above the $3.0$~TeV of Layout~A, because the injection term of Eq.~\eqref{eq:hybridrcs} does not scale with the circumference. Pulsed HTS dipoles of $\pm3$~T give $\sqrt{s}\approx8$~TeV. The assumptions of Neuffer and Shiltsev, $16$~T fixed and $\pm3.8$~T pulsed dipoles, give $E_\mu\approx5.2$~TeV and $\sqrt{s}\approx10$~TeV. Their packing factor of $0.85$ would raise these energies by $14$--$18\%$.

The collider ring of Layout~C has the circumference of the synchrotron and runs at its extraction energy. By Eq.~\eqref{eq:crcs} the two rings then have the same average field, $\bar B\approx3.2$, $4.0$, and $5.2$~T at the three points. The synchrotron therefore sets $\bar B$ for the collider. At $f=0.66$ (Section~\ref{subsec:energyB}) the collider needs arc dipoles of only $4.8$, $6.1$, and $7.9$~T, and the lowest point is close to the NbTi row of Table~\ref{tab:unk}. With the reference beam parameters scaled in energy as in the IMCC design, the luminosity scales as $\bar B\,E_\mu^2$ (Sections~\ref{sec:ring} and~\ref{subsec:unkreach}), and the target grows as $E_\mu^2$. The luminosity is thus a fraction $\bar B/9.2~\mathrm{T}\approx0.35$, $0.44$, and $0.57$ of the $(\sqrt{s}/10~\mathrm{TeV})^2$ target. The two upper points need pulsed magnets that have not been built. The survival relation of Ref.~\cite{NeufferShiltsev2018} links the ramp to the muon survival. In the stage from $1.5$ to $5.2$~TeV in the UNK tunnel, a survival of $82\%$, as in the last synchrotron of Ref.~\cite[Table~14.3.3]{IMCC2025}, needs about $21$~GeV per turn. The ramp then lasts $12$~ms, and the pulsed dipoles ramp at $\approx600$~T/s. Neuffer and Shiltsev accept $50\%$ survival per stage. This lowers the ramp rate to $\approx180$~T/s but reduces the luminosity, which is proportional to $N^2$, by a factor $(0.50/0.82)^2\approx0.37$. They expect rates of $\sim\!1000$~T/s to be achievable in HTS magnets~\cite{NeufferShiltsev2018}, whereas the fastest HTS test magnet has reached $\approx300$~T/s over a $0.5$~T span (Section~\ref{sec:acceleration}). All UNK numbers of this layout are estimates from Eqs.~\eqref{eq:crcs} and~\eqref{eq:hybridrcs} and from the survival relation.

Layout~C needs the fewest new ring tunnels. Only the first three synchrotrons remain to be housed, in about $17$~km of ring tunnel. In the IMCC chain the first two share a $6.0$~km tunnel, and the third has $10.7$~km (Section~\ref{sec:acceleration}). If the third synchrotron also moved into the UNK tunnel, as the second and third share the LHC tunnel at CERN, about $6$~km would remain. The UNK tunnel would then hold three rings, as its original design foresaw, but with very different magnets. The layout has its own difficulties. Its energy and its luminosity are both set by the pulsed magnets of the synchrotron. The $5.1$~m cross-section must hold the collider dipoles with their tungsten liners and the large-aperture hybrid synchrotron with its power converters. It must also hold the synchrotron RF, of which the last synchrotron of the IMCC greenfield chain alone has $2.8$~km~\cite[Table~I.8]{MuColParams2025}. The fast-ramping fields and eddy currents of the synchrotron would act next to the superconducting collider magnets, and their effect on the field quality and heat load of these magnets has not been studied. The six straight sections, $3.56$~km in total, must hold this RF together with the interaction region, the detector hall, injection, extraction, and collimation. Near the detector hall the synchrotron would need a bypass. As in Layout~B, these straight sections belong to the collider ring, and each emits a concentrated neutrino flux (Section~\ref{subsec:unkinfra}). The movers that deform the collider orbit (Section~\ref{sec:ring}) must work without disturbing the synchrotron in the same tunnel. No design study of this layout exists.

The three layouts trade energy, luminosity, and civil construction against one another (Table~\ref{tab:layouts}). Layout A reuses the tunnel for the item that needs the longest tunnel and the largest RF installation, and it keeps the collider ring short, which favours luminosity (Eq.~\eqref{eq:Lscaling}). Its energy with the IMCC baseline technology, about $6$~TeV, lies below the $10$~TeV reference. Layout B maximizes the energy, but it leaves the largest civil-engineering item, a tunnel $2.4$ to $3.5$ times longer than UNK, still to be built. Layout C needs the least new tunnel and only moderate collider dipoles, but its energy depends on pulsed magnets not yet demonstrated, and its luminosity stays below the target. All three layouts need the full front end and the lower acceleration stages on the site. Choosing among them is a first task for a design study. The next two subsections treat the energy reach and physics of Layout B, which the UNK circumference determines directly, and the physics of Layouts A and C in qualitative terms.

\begin{table}[!hbt]
\centering
\caption{The three layouts for the UNK tunnel (Section~\ref{subsec:layouts}). The new ring tunnels comprise the synchrotron and collider rings only, with $\approx17$~km for the first three IMCC synchrotrons. The front end, linacs, recirculating linacs, transfer lines, and halls are not included. The luminosity fraction refers to the $(\sqrt{s}/10~\mathrm{TeV})^2$ target at the reference beam parameters.}
\label{tab:layouts}
\small
\begin{tabular}{@{}c>{\raggedright\arraybackslash}p{2.4cm}>{\raggedright\arraybackslash}p{3.0cm}c>{\raggedright\arraybackslash}p{5.9cm}@{}}\toprule
Layout & UNK tunnel hosts & New ring tunnels (km) & $\sqrt{s}$ (TeV) & Main limitation \\
\midrule
A & last synchrotron & $\approx24$--$25$, including a $7$--$9$~km collider ring, or $\approx13$--$16$ with two synchrotrons in the UNK tunnel & $\approx6$ & Energy set by the fast-ramping magnets of the synchrotron. RF straight sections in a shallow tunnel concentrate the neutrino flux. \\
\addlinespace
B & collider ring & $\approx67$--$90$, including a $50$--$73$~km last synchrotron & $\approx14$--$21$ & Longest new tunnel and $11$--$16$~T collider dipoles. The fixed straight sections of the collider emit concentrated neutrino fluxes. \\
\addlinespace
C & collider ring and last synchrotron & $\approx17$, or $\approx6$ with the third synchrotron also in the UNK tunnel & $\approx6$--$10$ & Pulsed magnets above $6$~TeV not demonstrated. Luminosity $0.35$--$0.57$ of the target. Shared cross-section and straight sections. \\
\bottomrule
\end{tabular}
\end{table}

\subsection{Energy reach of Layout B}
\label{subsec:energyB}

\begin{figure}[!htb]
\centering
\includegraphics[width=0.66\linewidth]{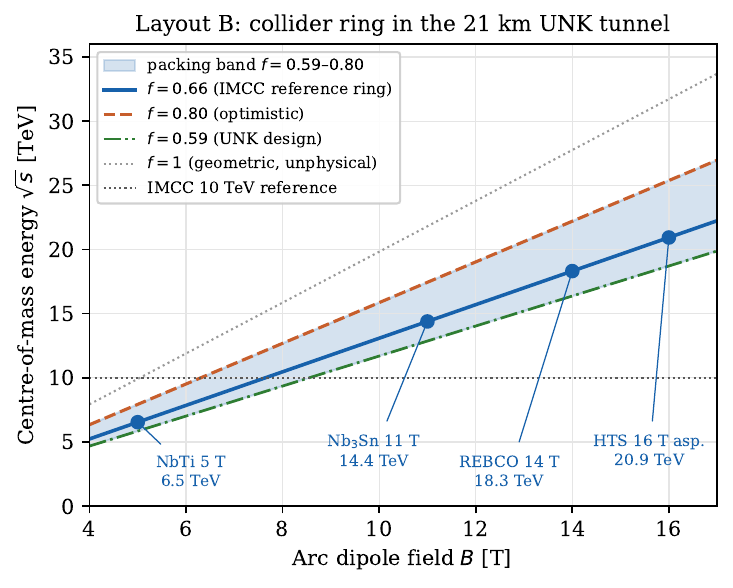}
\caption{Centre-of-mass energy of Layout~B, a muon collider ring in the 21~km UNK tunnel, as a function of the arc-dipole field, from the magnetic-rigidity relation. The shaded band spans dipole packing factors $f=0.59$--$0.80$ (UNK design to optimistic), and $f=1$ is the unphysical geometric bound. Markers indicate the four dipole options of Table~\ref{tab:unk}, and the horizontal line marks the IMCC 10~TeV reference. The collider of Layout~C lies on the same $f=0.66$ curve at $4.8$--$7.9$~T (Section~\ref{subsec:layouts}).}
\label{fig:unk}
\end{figure}

With the collider ring in the UNK tunnel, Eq.~\eqref{eq:rigidity} sets the energy. The geometric radius is $R = C/2\pi \approx 3306$~m, and the bending radius is $\rho = f\,R$. The dipole packing factor $f$ is the fraction of the circumference occupied by bending magnets, the remainder holding the interaction regions, the RF, injection, and the neutrino-mitigation hardware. Hence $\sqrt{s}\,[\mathrm{TeV}] = 2\times0.2998\,B\,[\mathrm{T}]\,f\,R\,[\mathrm{km}]$. The IMCC reference ring ($B=14$~T, $C=11.4$~km, hence $R=1814$~m) corresponds to $f\approx0.66$, the value adopted here~\cite{MuColParams2025}. The UNK design itself gives a lower calibration: its superconducting second stage was to store 3~TeV protons with 5.1~T dipoles~\cite{GurovUNK1995}, i.e.\ $f\approx0.59$. An optimistic value is $f=0.80$, close to the bound of $0.83$ set by the straight sections. Over $f=0.59$--$0.80$ the energies of Table~\ref{tab:unk} change by $-11\%$ to $+21\%$ (band in Fig.~\ref{fig:unk}).

\begin{table}[!hbt]
\centering
\caption{Centre-of-mass energy of Layout~B, the collider ring in the 21~km UNK tunnel ($R\approx3306$~m), from Eq.~\eqref{eq:rigidity} with $f=0.66$. The $5$, $11$, and $14$~T fields are those the IMCC considers achievable in the $138$--$158$~mm apertures required by the decay-heat liner~\cite{IMCC2025}. The IMCC baseline is $14$~T, and $16$~T is still assumed in the current MuCol lattice~\cite{MuColParams2025}.}
\label{tab:unk}
\begin{tabular}{lc>{\raggedright\arraybackslash}p{7.6cm}}\toprule
$B$ (T) & $\sqrt{s}$ (TeV) & Magnet technology \\
\midrule
5.0  & 6.5  & NbTi, $4.5$~K, $158$~mm (LHC-class technology) \\
11.0 & 14.4 & Nb$_3$Sn, $4.5$~K, $158$~mm (IMCC target) \\
14.0 & 18.3 & REBCO, $20$~K, $138$~mm (IMCC baseline) \\
16.0 & 20.9 & HTS, aspirational (current MuCol lattice) \\
\bottomrule
\end{tabular}
\end{table}

The usable field depends on the chosen technology and the coil aperture of the magnets. In the $\ge158$~mm aperture required by the decay-heat liner, NbTi is limited to $\approx5$~T~\cite{IMCC2025}, so LHC-class NbTi technology would give only $\sqrt{s}\approx6.5$~TeV in the UNK ring. Nb$_3$Sn dipoles at the IMCC target of $11$~T give $\sqrt{s}\approx14.4$~TeV, already above the IMCC $10$~TeV reference. REBCO dipoles at the $14$~T IMCC baseline give $\approx18.3$~TeV, and aspirational $16$~T dipoles $\approx20.9$~TeV. With the UNK design value $f=0.59$ each energy would be $11\%$ lower. With the dipoles that the IMCC is developing, the $21$~km circumference therefore places a muon collider ring above the reference energy, provided that the longer synchrotron tunnel of Section~\ref{subsec:layouts} is built.

\subsection{Projected physics reach}
\label{subsec:unkreach}

Because the muon-collider physics reach depends on the energy, all three layouts can in principle be placed on the existing reach curves. This subsection does so quantitatively for Layout~B only, and Table~\ref{tab:unkreach} refers to Layout~B. The curves assume the design luminosity at each energy, which any layout would have to be shown to deliver.

Layout A, at $\sqrt{s}\approx6$~TeV, lies between the IMCC $3$ and $10$~TeV stages of Section~\ref{sec:physics}. No dedicated projections exist at this energy, and none are interpolated here. Its kinematic limit for pair production is $\sqrt{s}/2\approx3$~TeV. With $14$~T dipoles its short collider ring keeps the luminosity lifetime of the reference ring, and with $11$~T dipoles about $80\%$ of it. The luminosity scaling of Section~\ref{sec:physics} sets its integrated-luminosity target at about $3.6$~ab$^{-1}$.

Layout C, at $\sqrt{s}\approx6$--$10$~TeV, also lies between the IMCC $3$ and $10$~TeV stages and reaches the $10$~TeV reference only at its upper point. Its pair-production limit is $\sqrt{s}/2\approx3$--$5$~TeV. Its luminosity is $35$--$57\%$ of the target (Section~\ref{subsec:layouts}). Its rate-limited measurements would therefore fall short of the projections at the same energy, while its kinematic limits would not change. No projections are interpolated here for this layout either.

For Layout B the luminosity caveat can be quantified at the scaling level. For the collider ring alone, the result favours the higher energies. Equation~\eqref{eq:Lscaling} gives the penalty of a fixed large ring. At $\sqrt{s}=10$~TeV the $20.8$~km UNK ring would deliver a fraction $C_{\rm ref}/C_{\rm UNK}=11.4/20.8\approx0.55$ of the luminosity of the $11.4$~km reference ring. Since the deficit is set by $\bar B$ (Eq.~\eqref{eq:nturnsB}), it closes when the ring is run at higher field. As far as the luminosity of the ring is concerned, the natural operating point of a fixed 21~km ring is therefore the \emph{top} of its energy range. The synchrotron tunnel and its RF, the neutrino dose and the required $\beta^*$ all become more demanding as the energy grows. With REBCO dipoles at the $14$~T IMCC baseline and $f=0.66$ ($\sqrt{s}\approx18$~TeV), the average bending field is $\bar B\approx9.2$~T. This equals the field of the $11.4$~km, 10~TeV reference, and so does the number of crossings per fill. Aspirational $16$~T dipoles ($\sqrt{s}\approx21$~TeV, $\bar B\approx10.6$~T) would give about $15\%$ more. Meeting the full $(\sqrt{s}/10~\mathrm{TeV})^2$ luminosity \emph{target} of the reference programme~\cite{BFW2021,MuColParams2025} would require more. It would need the reference beam brightness and bunch intensity. It would also need an interaction-point focusing that tightens as $\beta^*\propto1/E$~\cite{IMCC2025}, from $\beta^*=1.5$~mm at 10~TeV to $\beta^*\approx0.7$~mm at 21~TeV. No lattice with sufficient momentum acceptance exists yet even for $\beta^*=1.5$~mm (Section~\ref{sec:openchallenges}), and chromatic effects grow with energy~\cite{MuColParams2025}. Finally, several bunches per beam would be needed to avoid the loss in the decay-averaging factor once $f_r\gamma\tau_\mu\gtrsim1$~\cite{MuColParams2025}. At $\sqrt{s}=21$~TeV and $f_r=5$~Hz this factor is $f_r\gamma\tau_\mu\approx1.1$.

A Layout~B machine at $\sqrt{s}\approx14$--$21$~TeV sits between the computed $14$ and $30$~TeV points. It would measure the trilinear coupling $\lambda_3$ at roughly the $1.7$--$2.4\%$ level~\cite{DeBlas3TeV,IMCC2023}. Its direct pair-production reach would be $\sqrt{s}/2\approx7.2$--$10.5$~TeV, with VBF single production extending towards $m\approx14$--$21$~TeV~\cite{MuonSmasher2021,LFV2025}. The thermal wino and higgsino masses would remain within reach, subject to the luminosity caveat above~\cite{Capdevilla2021tracks,Capdevilla2024soft}. Several phenomenological studies of extra-dimensional gravitons, anomalous quartic gauge couplings, and axion-like particles have adopted $14$~TeV as a working benchmark, a value originally motivated by the LHC-tunnel reuse scenario~\cite{NeufferShiltsev2018,Delahaye2019}. They find sensitivities orders of magnitude beyond the HL-LHC~\cite{InanKisselevRS2023,Amarkhail2023gamma,Amarkhail2024Zgamma,InanKisselevALP2022}. That benchmark lies close to the lower end of the Layout~B range, and part of this work has been carried out at IHEP in Protvino (part of the NRC ``Kurchatov Institute''). Towards the upper end of the range, and at the assumed integrated luminosity, the reach exceeds the IMCC reference in every row of Table~\ref{tab:unkreach} except the thermal-WIMP row, where both cover the targets. The three UNK columns of the table are the Nb$_3$Sn ($11$~T), REBCO ($14$~T), and aspirational HTS ($16$~T) options of Table~\ref{tab:unk}, at the same $f=0.66$.

Each entry of Table~\ref{tab:unkreach} is an interpolation of the published energy-dependent reach reviewed in Section~\ref{sec:physics}, not the output of a dedicated UNK study. All UNK columns assume the integrated luminosity of the reference-programme target, $\int\!\mathcal{L}\,dt = 10~\mathrm{ab}^{-1}\times(\sqrt{s}/10~\mathrm{TeV})^2$~\cite{BFW2021,MuColParams2025}, given in the second row. The trilinear precision $\delta\lambda_3$ (68\% C.L.) is interpolated logarithmically between the values $3.6\%$, $2.5\%$, and $1.2\%$ computed at 10, 14, and 30~TeV (Fig.~\ref{fig:lambda3})~\cite{DeBlas3TeV,IMCC2023}. The first full-simulation analysis at 10~TeV, with $HH\to b\bar bb\bar b$ alone, gives about $7\%$ for the same luminosity~\cite{Andreetto2026}. The $\delta\lambda_3$ row therefore rests on the phenomenological projections, which are about twice as precise as the first full simulation. The yield $N_{HHH}=\sigma(\mu^+\mu^-\!\to\!\bar\nu\nu HHH)\times\int\!\mathcal{L}\,dt$ counts the Standard Model triple-Higgs events that give access to the quartic coupling $\lambda_4$. The cross sections are $\sigma=4.18$, $7.02$, and $18.5$~ab at 10, 14, and 30~TeV~\cite[Table~2]{Chiesa2020quartic}, interpolated logarithmically in between. Most of the growth comes from $M_{HHH}>3$~TeV, where the sensitivity to $\lambda_4$ is weak~\cite{Chiesa2020quartic}. The width-independent bound on the $\mu\mu H$ vertex, $\delta y_\mu$, is $\approx10\%$ at 10~TeV (95\% C.L., Higgs effective field theory, HEFT), or $\approx5\%$ in the dimension-6 SMEFT from $\mu^+\mu^-\to3H$~\cite{HiggsMuon2024}. No study exists above 10~TeV, so the UNK entries scale the HEFT value by $(\int\!\mathcal L\,dt/10~\mathrm{ab}^{-1})^{-1/2}$. This is conservative, because the sensitivity also grows with energy, from $25\%$ to $5\%$ between 3 and 10~TeV in the SMEFT case~\cite{HiggsMuon2024}. The direct pair-production reach is the kinematic limit $\sqrt{s}/2$. The VBF single-production reach is the kinematic ceiling $m\lesssim\sqrt{s}$ for a singly produced heavy state, whose mass is balanced by the forward beam remnants rather than by a second heavy particle~\cite{MuonSmasher2021,LFV2025}. The thermal-WIMP row marks which electroweak multiplets have their relic-abundance mass target covered~\cite{Capdevilla2021tracks,Capdevilla2024soft,Franceschini2023}. For the higgsino this rests on soft tracks at the 3~TeV stage~\cite{Capdevilla2024soft} and on disappearing tracks at 10~TeV ($5\sigma$ at $1.1$~TeV)~\cite{Capdevilla2021tracks}, neither yet studied at 14--21~TeV. The indirect scale $\Lambda$ is the SMEFT sensitivity, $30$--$100$~TeV at the 10~TeV reference and growing as $\sqrt{s}$~\cite{IMCC2025,EWscattering2024,HiggsMuon2024}. All values are indicative, statistics-driven estimates rounded to the quoted precision.

The fixed-ring penalty of Eq.~\eqref{eq:Lscaling} does not act on all rows of the table in the same way. Rate-limited quantities depend directly on the integrated luminosity: the statistical precision of $\delta\lambda_3$ and $\delta y_\mu$ scales as $1/\sqrt{\mathcal L}$ and the yield $N_{HHH}$ as $\mathcal L$. Kinematically limited quantities, the direct pair reach $\sqrt{s}/2$ and the VBF ceiling $m\lesssim\sqrt{s}$, are fixed by the beam energy. They do not depend on the luminosity as long as the signal is not limited by statistics, and a low luminosity degrades only how closely the ceiling is approached. Thermal-WIMP coverage lies in between: the target mass is fixed, but the significance at that mass depends on the number of signal events. The indirect scale $\Lambda$, derived from interference-dominated precision measurements, varies only as $\mathcal L^{1/4}$. At fixed bunch intensity, emittances, energy spread and repetition rate, and with $\beta^*=\sigma_z$, the luminosity scales as $\bar{\mathcal L}\propto\bar B\,E_\mu^2$~\cite[Eq.~(13)]{IMCC2023}, while the assumed target grows as $E_\mu^2$. A UNK ring with the reference beam parameters would deliver a fraction $\bar B_{\rm UNK}/\bar B_{\rm ref}\approx0.55\times\sqrt{s}/(10~\mathrm{TeV})$ of the luminosities in the second row. This fraction is $0.79$, $1.00$, and $1.15$ at $14.4$, $18.3$, and $20.9$~TeV, with several bunches per beam (see above). These are upper bounds. They assume that the interaction-region optics, the reference beam brightness, and the removal of the decay heat are all achieved. The rate-limited precisions would then change by factors of $1.12$, $1.00$ and $0.93$, and $\Lambda$ by less than $10\%$. The factor $0.55$, equivalent to a precision loss of $1/\sqrt{0.55}\approx1.35$, would apply only to a 10~TeV machine in the UNK ring. The heat load scales as $E_\mu/C$. In this ring it would range from $\approx250$~W/m at $10$~TeV to $\approx500$~W/m at $21$~TeV (an estimate by this scaling from the values of Section~\ref{sec:openchallenges}). The rate-limited entries of Table~\ref{tab:unkreach} are therefore conditional on a lattice and parameter study for this ring.

\begin{table}[!hbt]
\centering
\caption{Projected physics reach of Layout~B (collider ring in the UNK tunnel) versus the IMCC 10~TeV reference. The UNK columns use $f=0.66$, as in Table~\ref{tab:unk}.}
\label{tab:unkreach}
\setlength{\tabcolsep}{4.5pt}
\renewcommand{\arraystretch}{1.25}
\begin{tabular}{@{}>{\raggedright\arraybackslash}p{4.15cm}cccc@{}}\toprule
 & IMCC ref. & \multicolumn{3}{c}{Layout B (UNK collider ring)} \\
\cmidrule(lr){2-2}\cmidrule(lr){3-5}
Arc dipoles                 & $14$~T    & $11$~T Nb$_3$Sn & $14$~T REBCO & $16$~T HTS \\
$\sqrt{s}$ (TeV)            & 10        & 14.4 & 18.3 & 20.9 \\
Assumed $\int\!\mathcal{L}\,dt$ (ab$^{-1}$) & 10 & 21 & 34 & 44 \\
\midrule
Trilinear $\delta\lambda_3$ (68\% C.L.) & $3.6\%$ & $2.4\%$ & $1.9\%$ & $1.7\%$ \\
Triple-Higgs yield $N_{HHH}$ (SM, $\lambda_4$) & $\sim\!42$ & $\sim\!150$ & $\sim\!330$ & $\sim\!510$ \\
Muon Yukawa $\delta y_\mu$ (multi-boson, 95\% C.L.) & $\sim\!10\%$ & $\sim\!7\%$ & $\sim\!5.5\%$ & $\sim\!5\%$ \\
Direct pair reach ($m\!\le\!\sqrt{s}/2$) (TeV) & $5.0$ & $7.2$ & $9.2$ & $10.5$ \\
VBF single-production ceiling ($m\!\lesssim\!\sqrt{s}$) (TeV) & $\lesssim\!10$ & $\lesssim\!14$ & $\lesssim\!18$ & $\lesssim\!21$ \\
Thermal-WIMP targets covered (TeV) & \multicolumn{4}{c}{wino ($2.9$) $+$ higgsino ($1.1$)} \\
Indirect scale $\Lambda$ (SMEFT) (TeV) & $30$--$100$ & $45$--$145$ & $55$--$185$ & $65$--$210$ \\
\bottomrule
\end{tabular}
\end{table}

\subsection{Adapting the infrastructure}
\label{subsec:unkinfra}

Each layout would reuse the tunnel and its underground halls. Each would require a complete new injector complex on the site, as for any greenfield muon collider: the proton driver, target, cooling channel, and the lower stages of the acceleration chain. The acceleration chain is itself a large civil-construction item. In the IMCC greenfield design the last rapid-cycling synchrotron, which brings the beams to $5$~TeV, is $35$~km in circumference~\cite{MuColParams2025}. Layout~B would add ring-wide high-field dipoles and their cryogenics, a new interaction region with its tungsten shielding nozzles, and an excavation of the detector hall. It would also need a new tunnel of $50$--$73$~km for the last synchrotron (Section~\ref{subsec:layouts}). Layout~A would instead equip the UNK tunnel with the fixed and pulsed dipoles of the last synchrotron and with its RF, the largest part of the RF voltage of the chain~\cite[Table~I.8]{MuColParams2025}. It would add a new collider tunnel of $7$--$9$~km with its interaction region and detector hall. Layout~C would install both rings in the UNK tunnel, the collider with $5$--$8$~T dipoles and the hybrid synchrotron with its RF. It would need neither a separate collider tunnel nor a separate tunnel for the last synchrotron. According to IHEP, the tunnel has been maintained and is in good condition. The design study must establish whether the $5.1$~m diameter tunnel and the site services can accommodate HTS dipoles, their cryostats, and the heavy neutrino shielding (Layout~B). For Layout~A the same question applies to the RF and power converters of a synchrotron. For Layout~C both sets of equipment would have to share one cross-section. The IMCC cost estimate, made for the CERN option, is broken down by technical category (RF, magnets, power converters, civil engineering, and others) rather than by subsystem~\cite{IMCC2025}. A detailed cost split by system, including the share of the acceleration chain, is therefore not yet available. In that estimate the RF, including that of the cooling channel, weighs most at the low-energy stage, and the civil engineering is a modest share because the SPS and LHC tunnels are reused~\cite{IMCC2025}. The IMCC also notes that a significant part of the budget could potentially be saved by not having to build the costly and long synchrotron tunnels~\cite{IMCC2025}. Layout~A would realize this saving at Protvino, because the UNK tunnel houses the last synchrotron. Layout~B would save only the collider-ring tunnel. The new synchrotron tunnel of Layout~B, $50$--$73$~km, is longer than the $35$~km of the greenfield design. The case for Layout~B thus rests on its energy reach and not on a saving in civil construction. Layout~C would save both the tunnel of the last synchrotron and the collider-ring tunnel, at the price of energy and luminosity (Table~\ref{tab:layouts}).

The neutrino-flux constraint of Section~\ref{sec:ring} becomes more severe at higher energy. A $14$--$21$~TeV machine in Layout~B therefore demands \emph{more} aggressive mitigation than the $10$~TeV reference. The UNK tunnel depth of $20$--$60$~m (Section~\ref{subsec:unktunnel}) is, moreover, shallower than in the IMCC implementation at CERN, whose experimental shafts reach depths of $142$~m and $166$~m~\cite{IMCC2025}. For this depth the decay-plane neutrinos surface at $d\simeq\sqrt{2hR_\oplus}\approx16$--$28$~km from the ring, where they would reach populated districts of the Moscow and Kaluga regions. The shape of the tunnel matters as well. In Layouts~B and~C the six long straight sections, about $3.6$~km in total, are fixed by the existing tunnel, and each of them would emit a concentrated neutrino flux in a fixed direction. The IMCC design minimizes such straight sections outside the interaction region for this reason~\cite{IMCC2025}. In an existing tunnel their number, length, and orientation are set by the geometry, so each straight section would need a site-specific dose assessment. In Layout~A the collider ring is new and can be designed with the depth and short straight sections that the IMCC requires. The synchrotron in the UNK tunnel would also produce neutrinos. For the CERN option the IMCC notes that RF in the long straight sections of shallow existing tunnels could concentrate the flux, and that the resulting surface dose may prevent this placement~\cite{IMCC2025}. The UNK tunnel lies at only $20$--$60$~m, so this caveat applies to Layout~A at least as strongly. Layout~C combines both sources in the same straight sections. Its lower energy reduces the dose per decay. At $\sqrt{s}\approx6$~TeV the dose is about a third of the $10$~TeV value in the deformed arcs and about a sixth in the straight sections (an estimate by the scalings of Section~\ref{sec:ring}). In the CERN option the RF of the last synchrotron alone has about $2.9$~km of active cavity length~\cite{IMCC2025}. This is comparable to the $3.56$~km of all UNK straight sections, so the placement of the RF and the resulting flux must be part of the design study. The IMCC objective of $\approx10~\mu$Sv per year can be compared with the Russian norms. Russian law and radiation-safety norms set the public dose limit at $1$~mSv per year (averaged over any five consecutive years). They exempt sources that give rise to an individual annual effective dose of no more than $10~\mu$Sv~\cite{FZ3RadSafety1996,NRB2009}, so the IMCC objective coincides with the Russian exemption level. For the UNK option a site-specific radiological and legal study is a mandatory prerequisite of any decision on the project. It would have to cover the actual depth and geology of the tunnel and the population along the exit directions of the arcs and of the straight sections in the Moscow and Kaluga regions. It would also have to cover the deformation amplitude, land control along the exit sectors, and compliance with these norms~\cite{MokhovVanGinneken2000,IMCC2025}. This review regards the neutrino flux as the foremost technical caveat of the UNK option.

The same flux can also be put to use. Instead of only diluting the neutrinos, a straight section of the collider ring could send its collimated flux along a controlled direction that avoids populated districts. Ideally the flux would exit into unpopulated terrain or downward towards a dedicated underground hall. In Layouts~B and~C this direction is fixed by the existing straight sections, while the new ring of Layout~A could be oriented for the purpose. A muon storage ring viewed this way is a neutrino beamline. Extracting a well-characterized neutrino beam from the straight section of a muon storage ring goes back to Geer~\cite{Geer1998} and underlies the neutrino-factory and nuSTORM programmes~\cite{nuSTORM2025a,nuSTORM2025b} of Section~\ref{subsec:nu}. At collider energies the flux from a single straight section is far more energetic and forward-peaked, though less intense, than at a dedicated neutrino factory. A remote detector on that axis could therefore host a stand-alone high-energy neutrino experiment parasitic on collider operation. It would sit tens of kilometres downstream, where the beam has risen to the surface or been intercepted underground. The orbit deformation minimizes the dose but wastes the flux. An \emph{aimed} decay-plane section instead concentrates the flux into one azimuth, which must be depopulated and radiologically controlled over its full surfacing length, while the arcs are deformed as before. Whether the UNK site admits such a direction, given the $16$--$28$~km surfacing distance and the population of the Moscow and Kaluga regions, is a question for the same site-specific study.

\subsection{Context and caveats}

Under present conditions, with CERN--Russia cooperation suspended and participation in the IMCC institutionally closed to Russian laboratories, a UNK design study is realistic primarily as a broad national initiative. It would be positioned for reintegration into the international programme when circumstances allow. The tunnel is located on the territory of the A.~A.~Logunov Institute for High Energy Physics (IHEP) in Protvino. IHEP is now part of the NRC ``Kurchatov Institute'' (full designation NRC ``Kurchatov Institute''~--~IHEP). The NRC ``Kurchatov Institute'' plans to build the SILA synchrotron-radiation facility on the same site. The compatibility of the two projects and the rebuilding of the site's accelerator workforce belong to the design study as much as the tunnel itself.

A UNK-based muon collider would be a natural focus for the accelerator and particle-physics communities across the Russian institutes. These are JINR at Dubna, the Moscow institutes, the Budker Institute of Nuclear Physics at Novosibirsk, and the NRC ``Kurchatov Institute'', whose IHEP site at Protvino would host the tunnel. It would also be a concrete way for them to contribute to, and benefit from, the global muon-collider effort. The most valuable product of such a study would be the technology it drives: expertise in high-power targetry, six-dimensional ionization cooling, and high-field HTS magnets. This expertise is needed not only for a muon facility but also for other future high-power and high-field accelerators, and today it is concentrated in a few laboratories. With the 2026 European Strategy prioritizing FCC-ee while keeping muon-collider R\&D open as an alternative~\cite{ESPP2026}, a design study of a tunnel-reuse option is a contribution that fits the present phase of the programme. No conceptual design report exists for a UNK muon collider. The estimates here are therefore interpolations of the published energy-dependent reach and scalings of the IMCC parameters, intended to motivate a dedicated design study. The cryogenic and power infrastructure and the full cost would all have to be assessed in such a study. The maturation of the muon-beam technologies discussed in Section~\ref{sec:technology} is the item with the longest lead time.

\section{Non-collider applications of intense muon beams}
\label{sec:applications}

The muon source of a muon collider consists of a multi-megawatt proton driver, a high-field capture solenoid, and a cooling channel delivering bright, low-energy muon beams. Together with the stored beams it feeds, it is a scientific instrument in its own right. Five applications are discussed here: the neutrino programme, muon-catalyzed fusion, muography, muonic-atom isotopic analysis, and muon spin spectroscopy. They bear on the facility in different ways. The neutrino programme uses the stored muons themselves and belongs to the physics case of a muon-beam facility (Section~\ref{subsec:nu}). Muon-catalyzed fusion is limited today by the available negative-muon intensity and would gain directly from a collider-class source. Muography uses the natural cosmic-ray flux, while muonic-atom analysis and muon spin spectroscopy are served by existing dedicated muon sources that need only a small fraction of that intensity. For these three a collider front end would be an additional source, not a motivation for the facility. Each of the five is an active field, and Russian centres have contributed to several of them. At Dubna, the Dzhelepov Laboratory of Nuclear Problems of JINR measured muon-catalyzed fusion from the 1960s (Section~\ref{subsec:mucf}). It also confirmed the resonant absorption of negative muons by nuclei that had been predicted at Moscow State University (Section~\ref{subsec:muxray}). A group from the Institute of Theoretical and Experimental Physics (ITEP) in Moscow observed atomic muonium in solids by muon spin precession (Section~\ref{subsec:musr}). Ionization cooling, on which the bright beams of a collider front end depend, was proposed independently at the Institute of Nuclear Physics in Novosibirsk (now the Budker Institute) and at IHEP in Protvino~\cite{AdoBalbekov1971,Skrinsky1981}. Protvino also has a long record in accelerator neutrino physics (Section~\ref{subsec:nu}). The 1977 prediction that one muon can catalyze $\sim\!10^2$ fusions in a deuterium--tritium mixture came from IHEP and JINR (Section~\ref{subsec:mucf}). An early complex made of the proton driver, target, and cooling channel (the front end of Section~\ref{sec:technology}) would already supply beams for muon-catalyzed fusion, muonic-atom analysis, and, prospectively, active muography. A muon facility at Protvino could thus produce results well before the collider stage.

\subsection{The neutrino programme}
\label{subsec:nu}

Every muon decay yields two neutrinos, $\mu^-\to e^-\bar\nu_e\nu_\mu$ and its charge conjugate. A stored muon beam is therefore a neutrino source whose flavour content is fixed by the decay and whose flux follows from the stored current and the decay kinematics~\cite{Geer1998}. Three kinds of facility have been proposed to use this source, distinguished mainly by the energy of the stored muons. Neutrinos from Stored Muons (nuSTORM) would store muons of $\sim\!1$--$6$~GeV to deliver neutrino fluxes known to better than a percent. This would enable precision cross-section and electroweak measurements as well as searches for sterile neutrinos~\cite{nuSTORM2025a,nuSTORM2025b,EWnuE2025}. nuSTORM would have the highest stored muon-beam power of any facility so far~\cite{nuSTORM2025a,IMCC2025}. It could also supply the muon cooling demonstrator. The IMCC has considered a common target, with pions of about $200$~MeV for the demonstrator and of order a GeV for the neutrino source~\cite{IMCC2025}. A neutrino factory is a separate and larger facility. It would store muons below a few tens of GeV, aiming at $\sim\!10^{20}$ useful decays per year in long straight sections directed at a distant detector~\cite{IDSNF2011,Bogacz2022NF}. It would open the $\nu_e\to\nu_\mu$ and $\nu_e\to\nu_\tau$ oscillation channels, and with them tests of T and CPT invariance beyond the reach of DUNE and Hyper-Kamiokande~\cite{Bogacz2022NF,IMCC2025}. Its storage ring differs from that of a collider, but to leading order the two share the same front end of muon production, cooling, and acceleration~\cite{IMCC2025}. Protvino has used accelerator neutrino beams for decades. A U-70 beamline served the SKAT bubble chamber, the ITEP--IHEP spark-chamber spectrometer, and the IHEP--JINR neutrino detector~\cite{Akindinov2019}. For the future, the P2O proposal would send a beam from Protvino to the KM3NeT/ORCA detector over a $2595$~km baseline~\cite{Akindinov2019}.

The collider ring is the third source. At multi-TeV energies most muons decay in the arcs, and the usable neutrinos come from the straight sections. Their intensity is lower than at a neutrino factory and their energy too high for oscillation experiments. They would, however, give access to neutrino fixed-target scattering at energies never reached in the laboratory. For the 10~TeV reference the IMCC estimates $9\times10^{16}$ neutrinos of each species per year from a 10~m field-free region around the interaction point. The full straight section would give a few to ten times more. About $30\%$ of them would cross a target of 10~cm radius in a forward detector $200$~m from the interaction point~\cite{IMCC2025}. A straight section of the collider ring might instead be aimed so that its flux reaches a remote detector along a controlled direction (Section~\ref{subsec:unkinfra}). Fixed-target use of the degrading muon beam (muon on storage-ring target, $\mu$OST) would probe nucleon structure beyond the reach of the EIC~\cite{muOST2025,BeamDump2024,NuPortalDM2025}.

\subsection{Muon-catalyzed fusion}
\label{subsec:mucf}

A negative muon stopped in a deuterium--tritium mixture is captured into an atomic orbit around a single nucleus, forming a muonic atom (\dmuon\ or \tmuon). This atom is about $207$ times smaller than the corresponding electronic atom, because the muon is that much heavier than the electron. The muon settles preferentially on the triton, the more deeply bound of the two. This small, electrically neutral \tmuon\ atom then slips inside the electron cloud of a neighbouring hydrogen-isotope molecule and is captured by one of its deuterons. The result is the three-body muonic molecular ion $dt\mu$, in which a deuteron and a triton are bound together by the muon instead of the shared electrons. Because the muon draws the two nuclei to an internuclear separation some $207$ times smaller than in an ordinary molecule ($\sim\!500$~fm), they tunnel through the Coulomb barrier and fuse. The muon is then released and can catalyze further fusions. The effect was discovered experimentally by Alvarez and collaborators, who observed muon-catalyzed $pd$ fusion in a hydrogen bubble chamber~\cite{Alvarez1957} (after theoretical predictions by Frank and by Sakharov~\cite{Gershtein1990}). Much of its systematic development took place at the Dzhelepov Laboratory of Nuclear Problems at JINR in Dubna. There a group that included P.~F.~Yermolov made early measurements of muon-molecular formation rates and of muon capture in hydrogen in the 1960s and 1970s~\cite{Dzhelepov1964,Bystritskii1974}. Three-body calculations developed in Dubna in 1973--1977 gave the muonic-molecule energy levels with reasonable accuracy~\cite{Gershtein1990}. In 1977 Gershtein (IHEP) and Ponomarev (JINR) predicted that, because $dt\mu$ has a weakly bound state, one muon in a deuterium--tritium mixture can catalyze $\sim\!10^2$ fusions~\cite{GershteinPonomarev1977}. The classic review by Gershtein, Petrov and Ponomarev laid out the theory of the catalytic cycle and its central limitation, $\alpha$-sticking. Occasionally the muon released by a fusion is captured by the energetic $\alpha$ particle and lost from the cycle. This caps the number of fusions each muon can catalyze at roughly $150$, below the threshold for net energy production~\cite{Gershtein1990}. Interest has revived recently, with modern reaction-theory treatments~\cite{WuKamimura2024} and proposals to use muon-catalyzed fusion for targeted isotope production~\cite{ParisiIsotope2025}. Among the low-energy applications discussed here, muon-catalyzed fusion is the one limited by muon intensity. The most intense negative-muon beams today, at J-PARC and PSI, deliver $\sim\!(2$--$4)\times10^{7}~\mu^-/\mathrm{s}$, and Ref.~\cite{ParisiIsotope2025} concludes that even isotope production needs higher rates. The $10$~TeV collider design, by comparison, injects about $9\times10^{19}$ muons of each sign per year~\cite{IMCC2025}, of order $10^{13}$ per second over the assumed $10^{7}$~s of annual operation. A collider-class front end would therefore give these studies access to a new intensity range, provided that a dedicated line decelerates the negative muons and stops them in a dense hydrogen-isotope target. The $\alpha$-sticking limit on the energy balance is unaffected.

\subsection{Muography}

High-energy muons penetrate hundreds of metres of rock. The attenuation of the natural cosmic-ray muon flux, or prospectively of an accelerator muon beam, therefore images the density of large objects, much as an X-ray images a body. Cosmic-ray muography has already given results such as the discovery of a $30$~m void above the Grand Gallery of Khufu's Pyramid~\cite{Morishima2017} and the subsequent characterization of a corridor behind its north face~\cite{ScanPyramids2023}. It has also produced high-resolution density maps of active volcanoes for eruption monitoring~\cite{Olah2018,Saracino2024} and has been applied to nuclear-reactor imaging, for example at Fukushima Daiichi~\cite{Fujii2020}, and to civil-structure inspection. All of these were obtained with cosmic-ray muons and need no accelerator. An accelerator beam would make the method active. A directional muon beam of chosen energy would cross an object placed between the beam exit and a downstream tracking detector and map its density along each beam path. The exposures would be far shorter, and the angular resolution finer, than the weak cosmic flux of fixed spectrum allows. The candidate objects are those that can be brought to a beam line or lie close to one, such as large engineered structures, shielded containers, or the rock surrounding an underground hall of the facility. The beam would be a muon beam extracted from the front end, or from an early acceleration stage where greater penetration is needed. The collider ring and its neutrino flux, whose direction is fixed by the ring geometry, would play no role. This accelerator-based mode remains conceptual.

\subsection{Muonic-atom isotopic and elemental analysis}
\label{subsec:muxray}

When a negative muon cascades into the Coulomb field of a nucleus, it emits characteristic muonic X-rays at energies (around $6$~MeV for heavy elements) far above electronic transitions. These lines allow non-destructive, depth-selective elemental analysis, and their fine structure even distinguishes isotopes~\cite{Ninomiya2019}. The method rests on the physics of the muonic-atom cascade, established experimentally by Fitch and Rainwater in 1953~\cite{FitchRainwater1953}. A related contribution concerns the interaction of the muon with the nucleus itself. The resonant absorption of negative muons with excitation of collective nuclear levels was predicted by Balashov and Kabachnik at the Skobeltsyn Institute of Nuclear Physics of Moscow State University in 1963. It was confirmed by the Evseev group at the Dzhelepov Laboratory of JINR in 1968--1969 and entered as Discovery No.~173 in the USSR State Register in 1976~\cite{Discovery173}. The method is today applied to archaeology (provenancing of bronzes), materials science, and nuclear safeguards. The negative-muon beams it needs are available at existing dedicated sources (J-PARC and PSI, Section~\ref{subsec:mucf}), with a small fraction of the intensity a collider front end must produce. For this technique a collider programme would contribute cooled, low-emittance muon beams. They could extend the capabilities of the existing muon sources~\cite{IMCC2023} and, through a better-defined momentum, sharpen the depth selectivity of the analysis.

\subsection{Muon spin spectroscopy}
\label{subsec:musr}

Implanted positive muons are highly sensitive local magnetic probes: muon spin rotation, relaxation, and resonance (collectively $\mu$SR) is a standard tool in magnetism, superconductivity, battery-electrolyte diffusion, and chemical kinetics~\cite{Hillier2022}. In 1967, at the JINR synchrocyclotron in Dubna, an ITEP group observed the spin precession of atomic muonium in quartz, solid carbon dioxide, and ice~\cite{Myasishcheva1968}. The technique relies on intense, highly polarized low-energy muon beams, which dedicated facilities provide today, among them ISIS, J-PARC, PSI, and TRIUMF~\cite{IMCC2023,Hillier2022}. These beams carry only a small fraction of the muon flux that a collider front end must produce. A collider programme could add a $\mu$SR source only through a dedicated surface-muon extraction line with a suitable bunch time structure. Its more specific contribution would be technical. Low-energy polarized beams may benefit from cooling techniques related to ionization cooling, and the IMCC has established a collaboration with the ISIS muon group on the design of a low-energy muon cooling system~\cite{IMCC2025}.

\section{Discussion}
\label{sec:discussion}

The case for a muon collider rests on its physics reward, the maturity of its technology, and the deliverable luminosity. This review has tried to weigh each. The physics reward is large and in places unique. It includes sub-percent Higgs couplings, the muon Yukawa coupling, the trilinear and even quartic Higgs self-couplings, a near-complete closure of the thermal WIMP dark-matter window, and a broad direct-search programme. All of this comes at a parton-level energy comparable to that of a $100$~TeV proton collider but with clean lepton-collision final states. At present, the technology is not mature. Some ingredients have passed first milestones. MICE observed the reduction of transverse emittance by ionization cooling, in a single absorber and without re-acceleration, but did not demonstrate a practical cooling scheme. Small-bore laboratory solenoids have reached fields of the magnitude needed for final cooling, although not yet with the required bore. Most of the machine, however, rests on proposed solutions still to be demonstrated at the required performance. They include the proton compressor ring and the target inside a high-field solenoid, RF cavities at high gradient in strong magnetic fields, and the rapid-cycling synchrotrons. They also include the interaction-region magnets and the heat load from muon decays (Section~\ref{sec:openchallenges}). Detector studies indicate, in simulation, that the beam-induced background can be controlled. The IMCC itself notes that no interaction-region lattice yet has sufficient momentum acceptance, even for a perfect machine, and that its assumptions on large-aperture magnets have been somewhat optimistic. It also notes that the ring deformation for neutrino-dose mitigation requires magnet movements never seen in an accelerator~\cite{IMCC2025}. The design luminosity, on which the physics projections rest, is a target and has yet to be shown achievable. A cooling demonstrator, accelerator-quality high-field magnets, and a conceptual design are the next steps, on a timescale that the international roadmaps place at mid-century.

In this context the UNK tunnel offers an advantage that depends on the layout (Section~\ref{subsec:layouts}). Layout~A places the last synchrotron in the UNK tunnel, as the IMCC option at CERN does with the LHC tunnel. With the IMCC baseline technology it gives $\sqrt{s}\approx6$~TeV and needs a new collider ring of $7$--$9$~km. It reuses the tunnel for the part of the complex that needs the longest tunnel and the largest RF installation. The neutrino flux from the RF straight sections of a shallow tunnel is, however, a potential obstacle that the IMCC notes for the CERN option. Layout~B places the collider ring in the UNK tunnel. Nb$_3$Sn dipoles at the IMCC target field then give $\sqrt{s}\approx14$~TeV, REBCO dipoles at the $14$~T baseline $\approx18$~TeV, and aspirational $16$~T dipoles $\approx21$~TeV. At the upper end of this range the physics reach would exceed the IMCC reference, provided the luminosity can be delivered (Section~\ref{subsec:unkreach}). Layout~B, however, needs a new synchrotron tunnel of $50$--$73$~km, longer than the last synchrotron of the greenfield design. Layout~C places both the collider ring and the last synchrotron in the UNK tunnel, as Neuffer and Shiltsev proposed for the LHC. It needs the fewest new ring tunnels, about $6$--$17$~km, and collider dipoles of only $5$--$8$~T. Its energy is set by the pulsed magnets of the synchrotron. It is about $6$~TeV with the IMCC baseline magnets and about $10$~TeV only with $16$~T fixed and $\pm3.8$~T pulsed dipoles, which have not been demonstrated. Its luminosity is $35$--$57\%$ of the target. The principal risks are also specific. There is as yet no conceptual design report for any of the layouts, and the full injector complex and the high-field magnets would still have to be built. A design study must establish whether the $5.1$~m diameter tunnel and the site services can accommodate the magnets, cryogenics, RF, and neutrino shielding of any layout.

A final point concerns the applications of Section~\ref{sec:applications}. A muon collider would be reached through a sequence of increasingly bright muon sources, and the intermediate steps can already deliver science. The neutrino programme is the closest of them to the collider (Section~\ref{subsec:nu}). Muon-catalyzed fusion would profit from the intensity. Muography relies on cosmic rays, while muonic-atom analysis and condensed-matter spectroscopy are served by existing dedicated sources, so for these three the front end would be an added opportunity. Much of this work has a long history in the Russian institutes that would host and build such a facility.

\section{Conclusion}
\label{sec:conclusion}

This review has drawn together the physics case, the technological status, and the prospects for reusing the 21~km UNK tunnel for a muon collider. Three conclusions follow. First, the physics reach of a multi-TeV muon collider is broad and in places unique. It includes the Higgs self-couplings and the muon Yukawa coupling, a near-complete closure of the electroweak thermal-WIMP window, and energy-frontier reach in clean lepton collisions. It depends, however, on accelerator technologies that have not yet been demonstrated at the required performance and on a design luminosity that remains a target (Section~\ref{sec:openchallenges}). Second, the existing UNK tunnel is a quantifiable asset, but its best use is still open. As the tunnel of the last synchrotron it gives $\sqrt{s}\approx6$~TeV with the IMCC baseline technology and a new $7$--$9$~km collider ring. As the collider ring it gives $\sqrt{s}\approx14$--$18$~TeV for the $11$--$14$~T dipoles targeted by the IMCC, and up to $\approx21$~TeV with aspirational $16$~T dipoles. It then needs a separate $50$--$73$~km synchrotron tunnel. With both the collider ring and the last synchrotron in the tunnel, the facility needs the fewest new ring tunnels and only $5$--$8$~T collider dipoles. It then reaches about $6$~TeV with the IMCC baseline magnets and about $10$~TeV only with $16$~T fixed and $\pm3.8$~T pulsed synchrotron dipoles, neither yet demonstrated, at $35$--$57\%$ of the target luminosity. Third, the UNK estimates in this review scale the IMCC parameters, and these parameters themselves require much more detailed justification and technology development. The estimates are intended to motivate a dedicated design study. Only such a study can settle the layout of the acceleration chain, the deliverable luminosity, and, foremost, the neutrino-radiation mitigation.

Such a design study would serve two purposes. It would evaluate the UNK site itself, and it would build expertise in muon beams (targetry, ionization cooling, high-field magnets) that keeps its value wherever a muon collider is eventually built. That expertise, together with the science of the intermediate stages from neutrino physics to muon-catalyzed fusion, is a capability to which the Russian institutes have contributed for more than half a century. A UNK-based facility would let them contribute to it again.

\section*{Acknowledgements}
The study was conducted under the state assignment of Lomonosov Moscow State University.

%
%
%

\bibliographystyle{nsr}
\bibliography{references}

\end{document}